\documentclass[journal,twocolumn,twoside]{IEEEtranTCOM}

\normalsize

\usepackage{cite}

\usepackage[cmex10]{amsmath}
\usepackage{amssymb} 	% symbols
\usepackage{graphicx}

\usepackage{multirow}

\DeclareMathOperator{\Prob}{P\!}
\DeclareMathOperator{\Indicator}{I\!}
\DeclareMathOperator{\Qfunc}{Q\!}
\newcommand \Transfer[1]{\mathrm{T}_{#1}\!}
\def\dfree{d_\textnormal{free}}
\def\GF{\mathbf{GF}}
\def\SZ{\mathbf{S}^\textnormal{Z}}
\def\SD{\mathbf{S}^\textnormal{D}}
\def\SN{\mathbf{S}^\textnormal{N}}

\begin{document}
\title{Convolutional-Code-Specific CRC Code Design}

\author{Chung-Yu~Lou,~\IEEEmembership{Student~Member,~IEEE},
        Babak~Daneshrad,~\IEEEmembership{Member,~IEEE},
		and~Richard~D.~Wesel,~\IEEEmembership{Senior~Member,~IEEE}% <-this % stops a space
\thanks{C.-Y. Lou, B. Daneshrad, and R. D. Wesel are with the Department of Electrical Engineering, University of California, Los Angeles, CA 90095 USA (e-mail: chungyulou@ucla.edu; babak@ee.ucla.edu; wesel@ee.ucla.edu).}% <-this % stops a space
\thanks{This material is based upon work supported by the National Science Foundation under Grant Number 1162501. Any opinions findings, and conclusions or recommendations expressed in this material are those of the author(s) and do not necessarily reflect the views of the National Science Foundation.}}

%\markboth{IEEE Transactions on Communications}%
%{Lou \MakeLowercase{\textit{et al.}}: A New CRC Design for Convolutional Codes}

\maketitle

\begin{abstract}
%\boldmath
Cyclic redundancy check (CRC) codes check if a codeword is correctly received.  This paper presents an algorithm to design CRC codes that are optimized for  the code-specific error behavior of a specified feedforward convolutional code. The algorithm utilizes two distinct approaches to computing undetected error probability of a CRC code used with a specific convolutional code. The first approach enumerates the error patterns of the convolutional code and tests if each of them is detectable. The second approach reduces complexity significantly by exploiting the equivalence of the undetected error probability to the frame error rate of an equivalent catastrophic convolutional code. The error events of the equivalent convolutional code are exactly the undetectable errors for the original concatenation of CRC and convolutional codes.  This simplifies the computation because error patterns do not need to be individually checked for detectability.  As an example, we optimize CRC codes for a commonly used 64-state convolutional code for information length k=1024 demonstrating significant reduction in undetected error probability compared to the existing CRC codes with the same degrees.  For a fixed target undetected error probability, the optimized CRC codes typically require 2 fewer bits. 
\end{abstract}


% \fi
\IEEEpeerreviewmaketitle

\section{Introduction}
\IEEEPARstart{E}{rror}-detecting codes and error-correcting codes work together to guarantee a reliable link. The inner error-\emph{correcting} code tries to correct any errors caused by the channel. If the outer error-\emph{detecting} code detects any residual errors, then the receiver will declare a failed transmission. 

Undetected errors result when an erroneously decoded codeword of the inner code has a message that is a valid  codeword of the outer code.   This paper designs cyclic redundancy check (CRC) codes for a given feedforward convolutional code such that the undetected error probability is minimized.

\subsection{Background and Previous Work} %Otherwise, in order to guarantee a desired undetected error probability, a system may overshoot with a lower-rate error-detecting code than needed.  
A necessary condition for a good joint design of error-detecting and error-correcting codes, both using linear block codes, is provided in \cite{Error_Detection_Error_Correction_Guide}.  However, this condition is based on the minimum distances of the inner and outer codes and does not consider the detailed code structure.  

Most prior work on CRC design ignores the inner code structure by assuming that the CRC code is essentially operating on a binary symmetric channel (BSC).  We refer to this as the BSC assumption.  The BSC assumption does not take advantage of the fact that the CRC code will only encounter error sequences that are valid codewords of the inner code.

The undetected error probability of a CRC code under the BSC assumption was evaluated using the weight enumerator of its dual code in \cite{CRC_Undetected_Error_Dual_Code}. Fast algorithms to calculate dominant weight spectrum and undetected error probability of CRC codes under the BSC assumption were presented in \cite{CRC_Undetected_Error_Algorithm_1, CRC_Undetected_Error_Algorithm_2}. 

In \cite{Koopman_CRC_Length},  Koopman and Chakravarty list all standard and good (under the BSC assumption) CRC codes with up to $16$ parity bits for information lengths up to $2048$ bits. The authors recommend CRC codes given the specific target redundancy length and information length. 

For more than $16$ CRC parity bits, it is difficult to search all possibilities and find the best CRC codes even under the BSC assumption.  Some classes of CRC codes with $24$ and $32$ bits were investigated under the BSC assumption in \cite{CRC_24_32_Bit} and an exhaustive search for $32$-bit CRC codes under the BSC assumption was performed later in \cite{CRC_32_Bit}.  

Because all of these designs ignore the inner code by assuming a BSC, there is no guarantee of optimality when these CRC codes are used with a specific inner code.

A few papers do consider CRC and convolutional codes together. In \cite{CRC_Convolutional_Message_Length}, the CRC code was jointly decoded with the convolutional code and used to detect the message length without much degradation of its error detection capability. In \cite{CRC_Puncture}, CRC bits were punctured to reach higher code rates.  The authors noticed that bursty bit errors caused an impact on the performance of the punctured CRC code.

Recently, \cite{CRC_Correcting_Convolutional} and \cite{CRC_Correcting_Turbo} have considered a CRC code used for error correction jointly with convolutional and turbo codes, respectively. However, in these cases, the error detection capability of the CRC code is degraded. Such undetected error probability as well as false alarm probability were analyzed in \cite{CRC_Analysis_UEP_FAP} under the BSC assumption. The authors of \cite{CRC_Analysis_UEP_FAP} also modeled the bursty error at the turbo decoder output using a Gilbert-Elliott channel.

\subsection {Main Contributions}
We propose two methods to compute the undetected error probability of a CRC code concatenated with a feedforward convolutional code. The \emph{exclusion} method enumerates possible error patterns of the inner code and excludes them one by one if they are detectable. The \emph{construction} method constructs a new convolutional code whose error events correspond exactly to the undetectable error events of the original concatenation of CRC and convolutional codes.  With these two methods as tools, we design CRC codes for the most common 64-state convolutional code for information length $k=1024$ and compare with existing CRC codes, demonstrating the performance benefits of utilizing the inner code structure.

This paper is organized as follows: Section~\ref{sec:System_Model} provides the system model.  Section~\ref{sec:Undetectable_Analysis} presents the exclusion and construction methods for computing the undetected error probability of a CRC code concatenated with an inner convolutional code. Section~\ref{sec:Search_for_CRC} describes how these two methods can be used to design a CRC code to minimize the undetected error probability for a specific feedforward convolutional code and information length.  Section~\ref{sec:Numerical_Result} applies this design approach to the most common 64-state convolutional code for information length $k=1024$. Section~\ref{sec:Conclusion} concludes the paper.

\section{System Model}
\label{sec:System_Model}
\begin{figure}[!t]
\centering
\ifCLASSOPTIONonecolumn
\includegraphics[width=0.53\columnwidth]{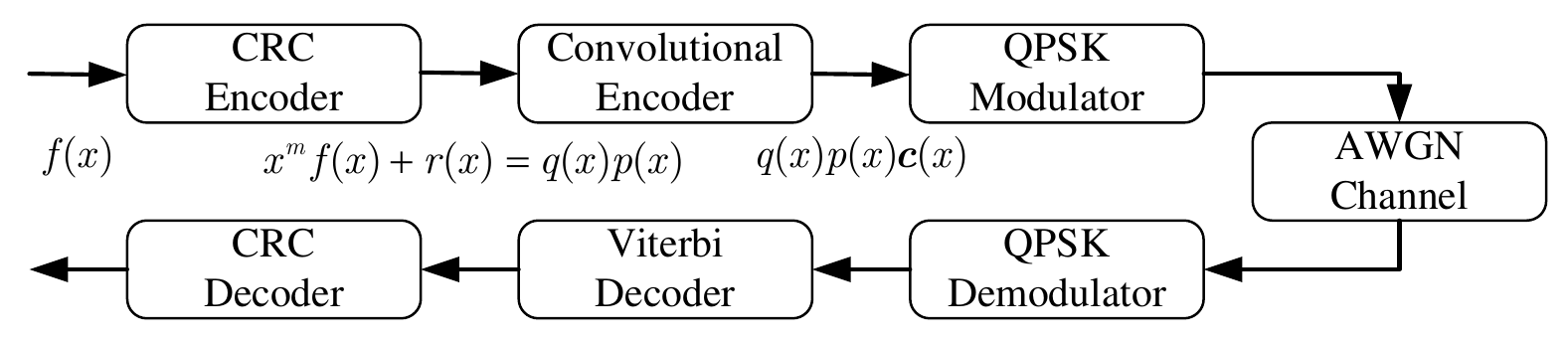}
\else
\includegraphics[width=0.99\columnwidth]{./figure/Tx_Rx_Block_small}
\fi
\caption{Block diagram of a system employing CRC and convolutional codes.}
\label{fig:Block_Diagram}
\end{figure}

Fig.~\ref{fig:Block_Diagram} shows the block diagram of our system model employing a CRC code concatenated with a convolutional code in an additive white Gaussian noise (AWGN) channel.
The $k$-bit information sequence is expressed as a binary polynomial $f(x)$   of degree smaller than or equal to $k-1$. 
%and determined by the information bits: the first bit determines the coefficient of $x^{k-1}$, the second bit determines the coefficient of $x^{k-2}$, and so on. At the CRC encoder output, $m$ redundant bits are appended to the information sequence. 
The $m$ parity bits are the remainder $r(x)$ of $x^mf(x)$ divided by the degree-$m$ CRC generator polynomial $p(x)$.  Thus the $n=k+m$-bit sequence described by $x^mf(x)+r(x)$ is divisible by $p(x)$ producing the $k$-bit quotient $q(x)$ and a remainder of zero.  The CRC-encoded sequence can also be expressed as $q(x)p(x)$. Thus $q(x)$ has a one-to-one relationship with $f(x)$.   Note that $q(x)p(x)$ is the result of processing the sequence $x^m q(x)$ ($q(x)$ and $m$ trailing zeros) by the CRC encoder circuit described by $p(x)$.  

The transmitter uses a feedforward, terminated, rate-$\frac{1}{N}$ convolutional code having $\nu$ memories with generator polynomial $\boldsymbol{c}(x)=\left[c_1(x), c_2(x), \cdots, c_N(x)\right]$.  The output $q(x)p(x)\boldsymbol{c}(x)$  of the convolutional encoder is sent to the AWGN channel using quadrature phase-shift keying (QPSK) modulation.  

Because the CRC bits are added at the end of the sequence, the highest degree term of $f(x)$ is the bit that is first in time and first to enter the convolutional encoder.  Consistent with this convention and in contrast to common representations, the highest degree terms of $\boldsymbol{c}(x)$ represent the most recent encoder input bits. Thus a convolutional encoder with the generator $G(D) = [1 + D^3 + D^4, 1 + D + D^2 ]$ will have  $\boldsymbol{c}(x)=[x^4 + x + 1, x^4 + x^3 + x^2]$.

The demodulated symbols are fed into a soft Viterbi decoder.  The CRC decoder checks the $n$-bit sequence resulting from Viterbi decoding. An undetected error occurs when the  receiver declares error-free decoding when the Viterbi decoder identified an incorrect codeword.

This work can be applied to feedback convolutional codes as well. Consider a feedback convolutional code with generator polynomial $\boldsymbol{c}(x)\slash c_\textnormal{FB}(x)$, where $c_\textnormal{FB}(x)$ is its feedback connection polynomial. This feedback code has the same set of codewords as a feedforward code with generator polynomial $\boldsymbol{c}(x)$. Assume the Viterbi decoder selects a wrong message $t(x)$ at the feedforward convolutional decoder output. In most cases, the same received signal is decoded as $t(x)c_\textnormal{FB}(x)$ by the feedback convolutional decoder. However, it is possible that the message decoded by the feedback decoder is not a multiple of $c_\textnormal{FB}(x)$. Such trellis deviation must happen during the termination of the codeword and corresponds to either a long error sequence with large codeword Hamming distance or a short error sequence occurring only at the end of the codeword. Either of these cases should not dominate the overall codeword error performance and thus we only consider message errors of the form $t(x)c_\textnormal{FB}(x)$.

% If $t(x)$ is divisible by $p(x)$, then $t(x)c_\textnormal{FB}(x)$ must be divisible by $p(x)$ and correct decoding is declared in both feedforward and feedback cases. If $t(x)$ is not divisible by $p(x)$, $t(x)c_\textnormal{FB}(x)$ could still be divisible by $p(x)$ when $p(x)$ and $c_\textnormal{FB}(x)$ are not relatively prime. Thus, for a given CRC, the undetected error rates need not be identical for feedforward and feedback encoders that have the same set of output codewords.

Let the greatest common divisor of $p(x)$ and $c_\textnormal{FB}(x)$ be $c_\textnormal{gcd}(x)$. The polynomial $p(x)$ divides $t(x)c_\textnormal{FB}(x)$ if and only if $p(x)\slash c_\textnormal{gcd}(x)$ divides $t(x)$. Hence, the undetected error probability of this CRC code concatenated with the feedback convolutional code is approximated by that obtained by the CRC code $p(x)\slash c_\textnormal{gcd}(x)$ concatenated with the feedforward convolutional code $\boldsymbol{c}(x)$ and can be analyzed using the methods presented in this paper. Furthermore, a smart choice when using a CRC code concatenated with a feedback convolutional code is to pick $p(x)$ and $c_\textnormal{FB}(x)$ relatively prime.

%For a system employing a tail-biting convolutional code, it is discussed at the end of Section~\ref{sec:Exclusive_Method}.

\section{Undetected Error Probability Analysis}
\label{sec:Undetectable_Analysis}
Let $e(x)$ be the polynomial of error bits in the Viterbi-decoded message so that the decoded $n$-bit sequence followed by $\nu$ zeros (for termination) is expressed as $q(x)p(x)x^{\nu}+e(x)$. If $e(x)\neq 0$ is divisible by $p(x)$, then this error is undetectable by the CRC decoder. This section presents two methods, the exclusion method and the construction method, to calculate the probability that a non-zero error $e(x)$ occurs that is undetectable.  

%This work focuses on the dominant error events of the convolutional code so that the analysis and the designed CRC codes are both most effective at high SNRs where overall behavior is well-characterized by these dominant events.
%Since the undetected error probability of a degree-$m$ CRC code on the BSC should approach $2^{-m}$ as channel error rate increases \cite{CRC_Undetected_Error_Upper_Bound}, the presented methods seek undetected error probabilities below that.

\subsection{Exclusion Method}
\label{sec:exclusion_Method}

The exclusion method enumerates the possible error patterns of the convolutional code and excludes the patterns detectable by the CRC code.  The probability of the unexcluded error patterns is the undetected error probability.   The exclusion method filters out part of the distance spectrum of the convolutional code through a divisibility test to create the distance spectrum of the undetectable errors of the concatenated code.

\subsubsection{Undetectable Single Error}
\label{sec:Exclusive_Single_Error}
An \emph{error event} occurs when the decoded trellis path leaves the encoded trellis path once and rejoins it once.  Let $e_{d,i}(x)$ be the polynomial of message error bits associated with the $i^\textnormal{th}$ error event that leads to a codeword distance $d$ from the transmitted codeword, where the range of $i$ is later specified in \eqref{eqn:PUD_Single_Approximation}. Note that in this paper, the term ``distance'' always refers to the convolutional code output Hamming distance. Let this error event have length $l_{d,i}$. Both $e_{d,i}(x)$ and $l_{d,i}$ are obtained through computer search of the given convolutional code. The first (highest power) term of $e_{d,i}(x)$ is $x^{l_{d,i}-1}$ and the last (lowest power) term is $x^\nu$ because every error event starts with a one and ends with a one followed by $\nu$ consecutive zeros. 

If the received data frame contains only one error event, then the polynomial of message error bits can be expressed as $e(x)=x^ge_{d,i}(x)$, where $g \in \left[0,n + \nu - l_{d,i}\right]$ indicates the possible locations where this error event may appear. If $e_{d,i}(x)$ is divisible by $p(x)$, this error event, including all of its offsets $g$, will be undetectable.  

The union bound of such error probability is given by
\ifCLASSOPTIONtwocolumn
\begin{IEEEeqnarray}{rCl}
\textnormal{P}_{\textnormal{UD},1}&\le&\sum_{d = \dfree}^\infty {\sum_{i = 1}^{a_d} {\Indicator\left(p(x)\mid\!e_{d,i}(x)\right)}}\nonumber\\
&&\max\left\{0, n + \nu - l_{d,i} + 1 \right\}\Prob\left( d \right)
\label{eqn:PUD_Single_Approximation},
\end{IEEEeqnarray}
\else
\begin{equation}
\textnormal{P}_{\textnormal{UD},1}\le\sum_{d = \dfree}^\infty {\sum_{i = 1}^{a_d} {\Indicator\left(p(x)\mid\!e_{d,i}(x)\right)}}\max\left\{0, n + \nu - l_{d,i} + 1 \right\}\Prob\left( d \right)
\label{eqn:PUD_Single_Approximation},
\end{equation}
\fi
where $\dfree$ is the free distance of the convolutional code, $a_d$ is the number of error events with output distance $d$, the indicator function $\Indicator\,(\cdot) $ returns one when $e_{d,i}(x)$ is divisible by $p(x)$ and zero otherwise, and $\Prob \left( d \right)$ is the pairwise error probability of an error event with distance $d$. The $\max$ operator ensures that the number of possible locations is always nonnegative even when $l_{d,i}$ is large. Note that while \eqref{eqn:PUD_Single_Approximation} does not explicitly use the generator polynomial of the convolutional code, it does implicitly depend on the generator polynomial because the generator polynomial determines the valid trellis error events $e_{d,i}(x)$. The subscript ``$1$'' in $\textnormal{P}_{\textnormal{UD},1}$ means that this probability only includes undetectable errors that are single error events. We call this type of error an \emph{undetectable single error}.  

For a QPSK system operated in an AWGN channel, $\Prob\,(d)$ can be computed using the tail probability function of standard normal distribution, i.e. Gaussian Q-function, as \cite{Viterbi_Textbook}
\begin{equation}
\Prob\left(d\right)=\Qfunc\left(\sqrt{2d\gamma}\right)\leq \Qfunc\left(\sqrt{2\dfree\gamma}\right)e^{-(d-\dfree)\gamma}
\label{eqn:Q_Approximation_Single},
\end{equation}
where $\gamma=\left.E_\textnormal{s}\middle/ N_0\right.$ is the signal-to-noise ratio (SNR) of a QPSK symbol, and $E_\textnormal{s}$ and $\left.N_0\middle/2\right.$ denote the received symbol energy and one-dimensional noise variance, respectively.   Note that the accuracy of \eqref{eqn:Q_Approximation_Single} comes in part from a knowledge of $\dfree$.  A useful Q-function approximation when knowledge of $\dfree$ is not available is presented in \cite{Convolutional_Code_PER_Milcom}. To generalize \eqref{eqn:Q_Approximation_Single} to a higher-order quadrature amplitude modulation (QAM) system with bit-interleaved coded modulation using a random interleaver, one can multiply $\gamma$ with a modulation-dependent factor to obtain an approximation. Details can be found in \cite{Convolutional_Code_PER_Milcom} as well.

To compute $\textnormal{P}_{\textnormal{UD},1}$ each error event $e_{d,i}$ must first be identified as either divisible by $p(x)$ or not.  One approach is to truncate \eqref{eqn:PUD_Single_Approximation} at $\tilde d$ to get an approximation, in which case all error events with distance $d \le \tilde d$ can be stored and this set of error events can be tested for divisibility by the CRC polynomial $p(x)$.  The choice of $\tilde d$ is based on the computational and storage capacity available to implement an efficient search such as \cite{Search_Distance_Spectrum}.  The required memory size to store the error events is proportional to $\sum_{d=\dfree}^{\tilde d} \sum_{i = 1}^{a_d} l_{d,i}$.  

The approximation of \eqref{eqn:PUD_Single_Approximation} can be quite tight if the probability of the terms with $d> \tilde{d}$ is negligible.  However, assuming that all error events with $d> \tilde{d}$ are undetectable provides
\ifCLASSOPTIONonecolumn
\begin{equation}
\textnormal{P}_{\textnormal{UD},1} \leq \sum_{d = \tilde d + 1}^{\infty} {n\,a_d\Prob\left( d \right)} + \sum_{d = \dfree}^{\tilde d} {\sum_{i = 1}^{a_d} {\left\{\Prob\left( d \right)\Indicator\left(p(x)\mid e_{d,i}(x)\right)\max \left\{ 0,n + \nu - l_{d,i}  + 1 \right\}\!\right\}}}
\label{eqn:PUD_Single_Bound},
\end{equation}
\else
\begin{IEEEeqnarray}{rCl}
\textnormal{P}_{\textnormal{UD},1} &\leq& \sum_{d = \tilde d + 1}^{\infty} {n\,a_d\Prob\left( d \right)} + \sum_{d = \dfree}^{\tilde d} {\sum_{i = 1}^{a_d} {\big\{\!\Prob\left( d \right)}}\nonumber\\
&&\cdot\Indicator\left(p(x)\mid e_{d,i}(x)\right)\max \left\{ 0,n + \nu - l_{d,i}  + 1 \right\}\!\!\big\}
\label{eqn:PUD_Single_Bound},
\end{IEEEeqnarray}
\fi
where $l_{d,i}$ for $d>\tilde d$ is replaced with $\nu+1$ because the shortest error event has length $\nu+1$. Note that \eqref{eqn:PUD_Single_Bound} can be computed because the error pattern $e_{d,i}(x)$ is only required for $d\leq\tilde d$, and the distance spectrum $a_d$ for $d>\tilde d$ provided by the transfer function  \cite{Viterbi_Convolutional_Code} can be used for the first term of \eqref{eqn:PUD_Single_Bound} as in the frame error rate (FER) bounds of \cite{Convolutional_Code_FER_LBA}.

In addition to the undetectable single error events discussed above, an undetectable error could consist of two or more error events, even though each of the error events itself is detectable. We will first discuss the case with two error events, and then generalize it to multiple error events.

\subsubsection{Undetectable Double Error}
\label{sec:Exclusive_Double_Error}
A double error involves two error events  $e_{d_1,i_1}(x)$ and  $e_{d_2,i_2}(x)$ with respective lengths $l_{d_1,i_1}$ and $l_{d_2,i_2}$.  To simplify notation, for $u \in \{1,2\}$ let $e_u(x)$ and $l_u$ refer to $e_{d_u,i_u}(x)$ and $l_{d_u,i_u}$, respectively. In a data frame with two error events, the polynomial of error bits in the message can be expressed as $e(x)=x^{g_1+g_2+l_2}e_1(x)+x^{g_2}e_2(x)$, where the exponents of two $x$'s tell the locations of the two error events. Furthermore, $g_1 \geq 0$ represents the interval of symbols (gap) between two error events and satisfies $g_1+g_2+l_1+l_2 \leq n + \nu$. If $x^{g_1 + l_2} e_1(x) + e_2(x)$ is divisible by $p(x)$, the error is an \emph{undetectable double error}.  Its length is $l_1+l_2+g_1$ and its offset is $g_2$. 

\begin{figure}[!t]
\centering
\ifCLASSOPTIONonecolumn
\includegraphics[width=0.53\columnwidth]{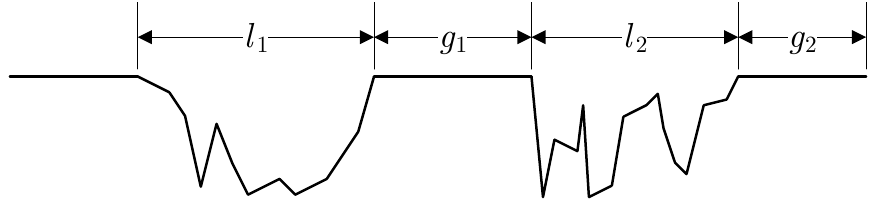}
\else
\includegraphics[width=0.99\columnwidth]{./figure/Double_Error_Illustration_small}
\fi
\caption{An illustration of two error events.}
\label{fig:Double_Error_Illustration}
\end{figure}

An upper bound of the probability of an undetectable double error occurring in the codeword is given by
\ifCLASSOPTIONonecolumn
\begin{IEEEeqnarray}{rCl}
\textnormal{P}_{\textnormal{UD},2} &\leq&\!\sum_{d_1=\dfree}^\infty {\:\sum_{d_2=\dfree}^\infty {\:\sum_{i_1 = 1}^{a_{d_1}} {\:\sum_{i_2 = 1}^{a_{d_2}} {\sum_{g_1 = 0}^{n + \nu - l_1 - l_2} {\Prob \left(d_1 + d_2\right)}}}}}\nonumber\\
&&\cdot\Indicator \left(p(x)\mid x^{g_1 + l_2} e_1(x) + e_2(x)\right)\left(n + \nu - l_1  - l_2  - g_1  + 1\right)
\label{eqn:PUD_Double_Bound1}.
\end{IEEEeqnarray}
\else
\begin{IEEEeqnarray}{rCl}
\textnormal{P}_{\textnormal{UD},2} &\leq&\!\sum_{d_1=\dfree}^\infty {\:\sum_{d_2=\dfree}^\infty {\:\sum_{i_1 = 1}^{a_{d_1}} {\:\sum_{i_2 = 1}^{a_{d_2}} {\sum_{g_1 = 0}^{n + \nu - l_1 - l_2} {\Prob \left(d_1 + d_2\right)}}}}}\nonumber\\
&&\cdot\Indicator \left(p(x)\mid x^{g_1 + l_2} e_1(x) + e_2(x)\right)\nonumber\\
&&\cdot\left(n + \nu - l_1  - l_2  - g_1  + 1\right)
\label{eqn:PUD_Double_Bound1} \, .
\end{IEEEeqnarray}
\fi
The distance of the double error event is simply the sum of the individual distances because the error events are completely separated as shown in Fig.~\ref{fig:Double_Error_Illustration}.
Two error events that overlap are simply a longer single error event, which was treated in Section~\ref{sec:Exclusive_Single_Error}.

Computation of \eqref{eqn:PUD_Double_Bound1} exactly is problematic because it requires infinite search depth.   Replacing the $l_1$ and $l_2$ of large-distance terms in \eqref{eqn:PUD_Double_Bound1} with $\nu+1$ yields a more computation-friendly upper bound similar to \eqref{eqn:PUD_Single_Bound} as follows:
\ifCLASSOPTIONonecolumn
\begin{IEEEeqnarray}{rCl}
\textnormal{P}_{\textnormal{UD},2} &\leq&\!\sum_{\left(d_1 ,d_2\right) \in \mathbf{D}_{\tilde d, 2}} {\:\sum_{i_1 = 1}^{a_{d_1}} {\:\sum_{i_2 = 1}^{a_{d_2}} {\sum_{g_1 = 0}^{n + \nu - l_1 - l_2} {\Prob \left(d_1 + d_2\right)
\Indicator \left(p(x)\mid x^{g_1 + l_2} e_1(x) + e_2(x)\right)}}}}\nonumber\\
&&\cdot\left(n + \nu - l_1  - l_2  - g_1  + 1\right)
+\!\sum_{\left(d_1 ,d_2\right) \notin \mathbf{D}_{\tilde d, 2}} {\frac{1}{2}\left(n - \nu\right)\left(n - \nu - 1\right)}
a_{d_1}a_{d_2}\Prob \left(d_1 + d_2\right)
\label{eqn:PUD_Double_Bound2},\IEEEeqnarraynumspace
\end{IEEEeqnarray}
\else
\begin{IEEEeqnarray}{rCl}
\textnormal{P}_{\textnormal{UD},2} &\leq&\!\sum_{\left(d_1 ,d_2\right) \in \mathbf{D}_{\tilde d, 2}} {\:\sum_{i_1 = 1}^{a_{d_1}} {\:\sum_{i_2 = 1}^{a_{d_2}} {\sum_{g_1 = 0}^{n + \nu - l_1 - l_2} {\Prob \left(d_1 + d_2\right)}}}}\nonumber\\
&&\cdot\Indicator \left(p(x)\mid x^{g_1 + l_2} e_1(x) + e_2(x)\right)\nonumber\\
&&\cdot\left(n + \nu - l_1  - l_2  - g_1  + 1\right)\nonumber\\
&&+\!\sum_{\left(d_1 ,d_2\right) \notin \mathbf{D}_{\tilde d, 2}} {\frac{1}{2}\left(n - \nu\right)\left(n - \nu - 1\right)}\nonumber\\
&&\cdot\;a_{d_1}a_{d_2}\Prob \left(d_1 + d_2\right)
\label{eqn:PUD_Double_Bound2},
\end{IEEEeqnarray}
\fi
where $\mathbf{D}_{\tilde d, 2} = \left\{\left(d_1 ,d_2\right) \left|d_1 ,d_2 \geq \dfree ,d_1 + d_2 \leq \tilde d\right.\right\}$. 

Because computational complexity limits the \emph{single} error event distance we can search, it is feasible to replace $\tilde{d}$ in $\mathbf{D}_{\tilde d, 2} $ with $\tilde{d}+ \dfree$. This is not particularly helpful because we have already assumed errors with distance $d>\tilde{d}$ are negligible or undetectable during the calculation of undetectable single errors.  We can also replace all $l_1$ and $l_2$ in \eqref{eqn:PUD_Double_Bound2} with $\nu+1$ and provide another upper bound which does not require any length information.

As described in Appendix~\ref{sec:Appendix_Search_Double_Error}, the number of $g_1$ values at which to check the divisibility of  $ x^{g_1 + l_2} e_1(x) + e_2(x)$ by $p(x)$ can be significantly reduced from $n+\nu-l_1-l_2+1$.
\subsubsection{Total Undetected Error Probability}
\label{sec:Exclusive_Undetected_Error_Probability}
In general, $s$ error events could possibly form an undetectable $s$-tuple error, whether each of them is detectable or not. The error bits in the message can be expressed as
\begin{equation}
e(x) = \sum_{u = 1}^s {\left(\prod_{v = u + 1}^s {x^{g_v  + l_v}} \right)x^{g_u} e_u(x)}.
\label{eqn:Composition_of_e(x)}
\end{equation}
This combined error will be undetectable if $e(x)$ is divisible by $p(x)$. Therefore, the probability of undetectable $s$-tuple errors $\textnormal{P}_{\textnormal{UD},s}$ can be approximated or bounded in the same way as \eqref{eqn:PUD_Double_Bound2}. For simpler notation, define sets
\begin{IEEEeqnarray*}{rCl}
\mathbf{D}_{\tilde d, s} &=& \left\{\!\left(d_1,\cdots,d_s \right)\left|d_u \geq \dfree\;\;\forall u, \sum_{u=1}^s {d_u} \leq \tilde d\right.\right\}\\
\mathbf{I}_s &=& \left\{\left(i_1,\cdots,i_s \right)\left| i_u \in \left[1,a_{d_u}\right]\;\;\forall u\in\left[1,s\right]\right.\right\}\\
\mathbf{G}_s &=& \left\{\!\left(g_1,\cdots,g_s \right)\left|g_u \geq 0\:\:\forall u, \sum_{u=1}^s {g_u}\leq n+\nu-\!\sum_{u=1}^{s+1}l_u\right.\!\right\}\!.
\end{IEEEeqnarray*}
Note that $\mathbf{G}_s$ is determined by all $d_u$ and $i_u$, and $\mathbf{I}_s$ is determined by all $d_u$.

Since an undetectable error may consist of any number of error events, the probability of having an undetectable error in the codeword $\textnormal{P}_\textnormal{UD}$ is upper bounded by $\sum_{s=1}^{\infty}{\textnormal{P}_{\textnormal{UD}, s}}$. Using the computation-friendly bound of each term such as \eqref{eqn:PUD_Single_Bound} and \eqref{eqn:PUD_Double_Bound2}, we obtain
\ifCLASSOPTIONonecolumn
\begin{IEEEeqnarray}{rCl}
\textnormal{P}_\textnormal{UD}\!&\leq\!&
\sum_{s = 2}^\infty {\Bigg\{\!\sum_{\left(d_1, \cdots ,d_s \right) \in \mathbf{D}_{\tilde d, s}} {\sum_{\left(i_1, \cdots ,i_s \right) \in \mathbf{I}_s} {\sum_{\left(g_1, \cdots ,g_{s - 1}\right) \in \mathbf{G}_{s - 1}} {\!\!\Indicator\left(p(x)\mid e(x)\right)
\left(n+\nu-\sum_{u=1}^s{l_u}-\sum_{u=1}^{s-1}{g_u}+1\right)}}}}\nonumber\\
&&\!\!\cdot\Prob\left(\sum_{u = 1}^s {d_u}\right)\!\!\!\Bigg\}
\!+\!\sum_{s=1}^\infty{\textnormal{P}_{>\tilde d,s}}
\!+\!\!\sum_{d_1 \in \mathbf{D}_{\tilde d,1}} {\!\!\left\{\!\Prob\left( d_1 \right)
\!\sum_{i_1 \in \mathbf{I}_1}{\!\Indicator\left(p(x)\mid e_1(x)\right)\max \left\{ 0,n + \nu - l_1  + 1 \right\}\!\!}\right\}}
%\textnormal{P}_\textnormal{UD}&\leq& \sum_{s=1}^\infty\textnormal{P}_{>\tilde d,s}
%+\!\sum_{d_1 \in \mathbf{D}_{\tilde d,1}} {\!\!\big\{\!\Prob\left( d_1 \right)}
%\sum_{i_1 \in \mathbf{I}_1}{\!\Indicator\left(p(x)\mid e_1(x)\right)\max \left\{ 0,n + \nu - l_1  + 1 \right\}\!\!\big\}}\nonumber\\
%&&+\sum_{s = 2}^\infty {\Bigg\{\sum_{\left(d_1, \cdots ,d_s \right) \in \mathbf{D}_{\tilde d, s}} {\!\Prob\left(\sum_{u = 1}^s {d_u}\right)}}
%\!\sum_{\left(i_1, \cdots ,i_s \right) \in \mathbf{I}_s} {\:\sum_{\left(g_1, \cdots ,g_{s - 1}\right) \in \mathbf{G}_{s - 1}} {\!\Indicator\left(p(x)\mid e(x)\right)}}\nonumber\\
%&&\cdot\left(n+\nu-\sum_{u=1}^s{l_u}-\sum_{u=1}^{s-1}{g_u}+1\right)\!\!\Bigg\}
\label{eqn:PUD_General_Bound},
\end{IEEEeqnarray}
\else
\begin{IEEEeqnarray}{rCl}
\textnormal{P}_\textnormal{UD}&\leq& \sum_{s=1}^\infty{\textnormal{P}_{>\tilde d,s}}+\sum_{s = 2}^\infty {\Bigg\{\sum_{\left(d_1, \cdots ,d_s \right) \in \mathbf{D}_{\tilde d, s}} {\!\Prob\left(\sum_{u = 1}^s {d_u}\right)}}\nonumber\\
&&\cdot\!\sum_{\left(i_1, \cdots ,i_s \right) \in \mathbf{I}_s} {\:\sum_{\left(g_1, \cdots ,g_{s - 1}\right) \in \mathbf{G}_{s - 1}} {\!\Indicator\left(p(x)\mid e(x)\right)}}\nonumber\\
&&\cdot\left(n+\nu-\sum_{u=1}^s{l_u}-\sum_{u=1}^{s-1}{g_u}+1\right)\!\!\Bigg\}+\!\sum_{d_1 \in \mathbf{D}_{\tilde d,1}} {\!\!\big\{\!\Prob\left( d_1 \right)}\nonumber\\
&&\cdot\sum_{i_1 \in \mathbf{I}_1}{\!\Indicator\left(p(x)\mid e_1(x)\right)\max \left\{ 0,n + \nu - l_1  + 1 \right\}\!\!\big\}}
%\textnormal{P}_\textnormal{UD}&\leq& \sum_{s = 1}^\infty {\Bigg\{\textnormal{P}_{>\tilde d,s}+\sum_{\left(d_1, \cdots ,d_s \right) \in \mathbf{D}_{\tilde d, s}} {\!\Prob\left(\sum_{u = 1}^s {d_u}\right)}}\nonumber\\
%&&\cdot\!\sum_{\left(i_1, \cdots ,i_s \right) \in \mathbf{I}_s} {\:\sum_{\left(g_1, \cdots ,g_{s - 1}\right) \in \mathbf{G}_{s - 1}} {\!\Indicator\left(p(x)\mid e(x)\right)}}\nonumber\\
%&&\cdot\max\left\{0,n+\nu-\sum_{u=1}^s{l_u}-\sum_{u=1}^{s-1}{g_u}+1\right\}\!\!\Bigg\}
\label{eqn:PUD_General_Bound},
\end{IEEEeqnarray}
\fi
where the composition of $e(x)$ depends on the number of error events $s$ as in \eqref{eqn:Composition_of_e(x)} and $\textnormal{P}_{>\tilde d,s}$ is the sum of probability of all $s$-tuple errors whose distances are greater than $\tilde d$. 

The probability sum of all large-distance $s$-tuple errors is
\begin{IEEEeqnarray}{c}
\textnormal{P}_{>\tilde d,s}=\!\!\!\sum_{\left(d_1, \cdots ,d_s \right) \notin \mathbf{D}_{\tilde d, s}}{\!\!\!\!\binom{n+\nu-s\nu}{s}\!\!\left(\prod_{u = 1}^s {\!a_{d_u}\!}\!\right)\!\Prob\left(\sum_{u = 1}^s {d_u}\!\right)}\!
\label{eqn:P_tilde_s_Definition},
\ifCLASSOPTIONtwocolumn
\IEEEeqnarraynumspace
\fi
\end{IEEEeqnarray}
where the combinatorial number represents the number of ways to have $s$ length-$(\nu+1)$ error events in a length-$(n+\nu)$ sequence. Using \eqref{eqn:Q_Approximation_Single} $\textnormal{P}_{>\tilde d,1}$ can be upper bounded by
\ifCLASSOPTIONonecolumn
\begin{equation}
\textnormal{P}_{>\tilde d,1}\leq\Qfunc\left(\sqrt{2\dfree\gamma}\right)e^{\dfree\gamma}\left\{\bar{\textnormal{P}}-\sum_{d_1 \in \mathbf{D}_{\tilde d, 1}}{n\,a_{d_1}e^{-d_1\gamma}}\right\}
\label{eqn:P_tilde_1_Bound},
\end{equation}
\else
\begin{IEEEeqnarray}{C}
\textnormal{P}_{>\tilde d,1}\leq\Qfunc\left(\sqrt{2\dfree\gamma}\right)e^{\dfree\gamma}\!\left\{\bar{\textnormal{P}}-\!\!\sum_{d_1 \in \mathbf{D}_{\tilde d, 1}}{\!\!n\,a_{d_1}e^{-d_1\gamma}}\right\}\!
\label{eqn:P_tilde_1_Bound},\IEEEeqnarraynumspace
\end{IEEEeqnarray}
\fi
where $\bar{\textnormal{P}}$ is defined as $\bar{\textnormal{P}}\triangleq \left.n\Transfer{}\left(D,L\right)\right|_{D=e^{-\gamma}\!,L=1}$ using the transfer function \cite{Viterbi_Convolutional_Code} 
\begin{equation}
\Transfer{}\left(D,L\right) = \sum_{d=\dfree}^{\infty} \sum_{i=1}^{a_d} D^d L^{l_{d,i}}\, .
\label{eqn:Transfer_Function}
\end{equation}
%\label{eqn:P_bar_Definition}.
%\end{equation}
Therefore, the sum of $\textnormal{P}_{>\tilde d, s}$ terms can be upper bounded by
\ifCLASSOPTIONonecolumn
\begin{IEEEeqnarray}{rCl}
\sum_{s = 1}^\infty \textnormal{P}_{>\tilde d,s}
&\leq&\sum_{s = 1}^\infty{\sum_{\left(d_1, \cdots ,d_s \right) \notin \mathbf{D}_{\tilde d, s}}{\frac{n^s}{s!}\left(\prod_{u = 1}^s {a_{d_u}}\right)\Prob\left(\sum_{u = 1}^s {d_u}\right)}}\nonumber\\
&\leq&\Qfunc\left(\sqrt{2\dfree\gamma}\right)e^{\dfree\gamma}\sum_{s = 1}^\infty{\left\{\frac{n^s}{s!}\sum_{\left(d_1, \cdots ,d_s \right) \notin \mathbf{D}_{\tilde d, s}}{\left(\prod_{u = 1}^s {a_{d_u}e^{-d_u\gamma}}\right)}\!\!\right\}}
\IEEEyessubnumber\label{eqn:P_tilde_Sum_Bound_1}\\
&=&\Qfunc\left(\sqrt{2\dfree\gamma}\right)e^{\dfree\gamma}\sum_{s = 1}^\infty{\left\{\frac{1}{s!}\bar{\textnormal{P}}^s-\frac{n^s}{s!}\sum_{\left(d_1, \cdots ,d_s \right) \in \mathbf{D}_{\tilde d, s}}{\left(\prod_{u = 1}^s {a_{d_u}e^{-d_u\gamma}}\right)}\!\!\right\}}
\IEEEyessubnumber\label{eqn:P_tilde_Sum_Bound_2}\\
&=&\Qfunc\left(\sqrt{2\dfree\gamma}\right)e^{\dfree\gamma}\left\{e^{\bar{\textnormal{P}}}-1
-\sum_{s = 1}^\infty\frac{n^s}{s!}\sum_{\left(d_1, \cdots ,d_s \right) \in \mathbf{D}_{\tilde d, s}}{\left(\prod_{u = 1}^s {a_{d_u}e^{-d_u\gamma}}\right)}\!\!\right\}
\IEEEyessubnumber\label{eqn:P_tilde_Sum_Bound_3}.
\end{IEEEeqnarray}
\else
\begin{IEEEeqnarray}{rCl}
\sum_{s = 1}^\infty \textnormal{P}_{>\tilde d,s}
&\leq&\sum_{s = 1}^\infty{\sum_{\left(d_1, \cdots ,d_s \right) \notin \mathbf{D}_{\tilde d, s}}{\frac{n^s}{s!}\left(\prod_{u = 1}^s {a_{d_u}}\right)\Prob\left(\sum_{u = 1}^s {d_u}\right)}}\nonumber\\
&\leq&\Qfunc\left(\sqrt{2\dfree\gamma}\right)e^{\dfree\gamma}\sum_{s = 1}^\infty{\Bigg\{\frac{n^s}{s!}}\nonumber\\
&&\cdot\sum_{\left(d_1, \cdots ,d_s \right) \notin \mathbf{D}_{\tilde d, s}}{\left(\prod_{u = 1}^s {a_{d_u}e^{-d_u\gamma}}\right)\!\!\Bigg\}}
\IEEEyessubnumber\label{eqn:P_tilde_Sum_Bound_1}\\
&=&\Qfunc\left(\sqrt{2\dfree\gamma}\right)e^{\dfree\gamma}\sum_{s = 1}^\infty{\Bigg\{\frac{1}{s!}\bar{\textnormal{P}}^s}\nonumber\\
&&-\:\frac{n^s}{s!}\sum_{\left(d_1, \cdots ,d_s \right) \in \mathbf{D}_{\tilde d, s}}{\left(\prod_{u = 1}^s {a_{d_u}e^{-d_u\gamma}}\right)}\!\Bigg\}
\IEEEyessubnumber\label{eqn:P_tilde_Sum_Bound_2}\\
&=&\Qfunc\left(\sqrt{2\dfree\gamma}\right)e^{\dfree\gamma}\Bigg\{e^{\bar{\textnormal{P}}}-1\nonumber\\
&&-\sum_{s = 1}^\infty\frac{n^s}{s!}\!\!\sum_{\left(d_1, \cdots ,d_s \right) \in \mathbf{D}_{\tilde d, s}}{\!\!\!\left(\prod_{u = 1}^s {a_{d_u}e^{-d_u\gamma}}\right)}\!\!\Bigg\}
\IEEEyessubnumber\label{eqn:P_tilde_Sum_Bound_3}.
\end{IEEEeqnarray}
\fi
The bound of Gaussian Q-function \eqref{eqn:Q_Approximation_Single} is used in \eqref{eqn:P_tilde_Sum_Bound_1}, and the transfer function is used to evaluate the sum of all $s$-tuple errors in \eqref{eqn:P_tilde_Sum_Bound_2}.  Using \eqref{eqn:PUD_General_Bound} and \eqref{eqn:P_tilde_Sum_Bound_3}, a bound of $\textnormal{P}_\textnormal{UD}$ can be calculated.  In fact, when an undetectable error occurs in a codeword, the receiver may still detect an error if a detectable error happens somewhere else in the codeword. Therefore, $\textnormal{P}_\textnormal{UD}$, which is the probability of having an undetectable error in the codeword, is an upper bound of the undetected error probability. When channel error rate is low, having a detected error occur along with the undetected error is a rare event so that our bound will be tight.

% If we were to consider a tail-biting convolutional code, only a minor modification is required. Although there are no terminating bits, the number of locations to start an error event is still $n$. Moreover, the number of available offsets of an undetectable error is $n$, which is independent of its length. Therefore, to obtain a bound for the undetected error probability, we just need to let the lengths of all undetectable errors be one and $\nu=0$, and the derivation from \eqref{eqn:PUD_General_Bound} to \eqref{eqn:P_tilde_Sum_Bound_3} will be valid.

Due to the limitation of searchable depth $\tilde d$ of error events, the exclusion method is only useful when undetectable errors with distances $d\leq \tilde d$ dominate. However, this requirement could be violated by a powerful CRC code that detects the short-distance errors.  The next subsection introduces the construction method, which allows the search depth to increase to distance $\hat{d}>\tilde{d}$.

\subsection{Construction Method}
\label{sec:Inclusive_Method}
The construction method utilizes the fact that all undetectable errors at the CRC decoder input are multiples of the CRC generator $p(x)$.  This method constructs an equivalent convolutional encoder  $\boldsymbol{c_{eq}}(x)=p(x)\boldsymbol{c}(x)$ to isolate these errors.   The set of non-zero codewords of $\boldsymbol{c_{eq}}(x)$ is exactly the set of erroneous codewords (given an all-zeros transmission) that lead to undetectable errors for the concatenation of the CRC generator polynomial and the original convolutional encoder.  Thus the probability of undetectable error is exactly the FER of $\boldsymbol{c_{eq}}(x)$. 

\begin{figure}[!t]
\centering
\ifCLASSOPTIONonecolumn
\includegraphics[width=0.53\columnwidth]{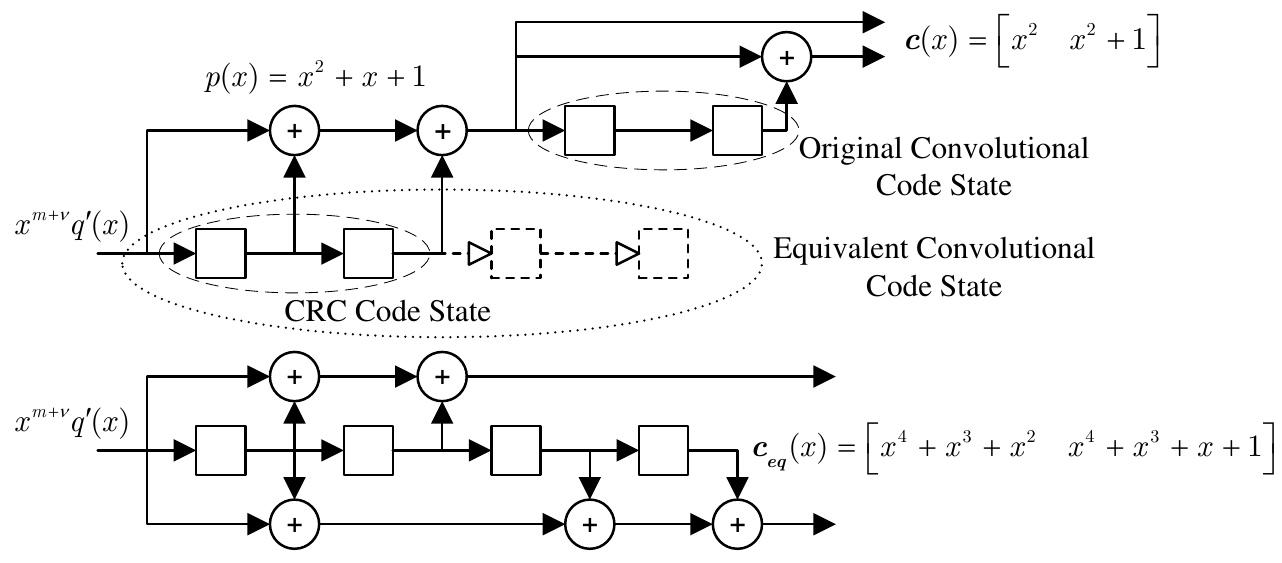}
\else
\includegraphics[width=0.99\columnwidth]{./figure/Equivalent_Encoder_small}
\fi
\caption{An example of CRC code, original convolutional code, and equivalent convolutional code.}
\label{fig:Equivalent_Encoder}
\end{figure}
Fig.~\ref{fig:Equivalent_Encoder} shows an example of how $\boldsymbol{c_{eq}}(x)$ is constructed from $\boldsymbol{c}(x)$ and $p(x)$, where $q'(x)$ with $m+\nu$ trailing zeros is the input that generates the undetectable erroneous codeword.  The input/output behavior of $\boldsymbol{c_{eq}}(x)$ is exactly the same as the concatenation of $p(x)$ and $\boldsymbol{c}(x)$.  For $p(x)$ with $m$ memory elements and $\boldsymbol{c}(x)$ with $\nu$ memory elements, $\boldsymbol{c_{eq}}(x)$ has $m+\nu$ memory elements.  At a given time, the state of the original encoder $\boldsymbol{c}(x)$ can be inferred from the state of $\boldsymbol{c_{eq}}(x)$ because the state of $\boldsymbol{c_{eq}}(x)$ contains the last $m+\nu$ inputs to $p(x)$ which are sufficient to compute the last $\nu$ outputs of $p(x)$, which exactly comprise the state of $\boldsymbol{c}(x)$.

\subsubsection{States of the Equivalent Encoder}
\label{sec:States_of_the_Equivalent_Encoder}
Define the all-zero state $\SZ$ to be the state where all memory elements of the equivalent encoder are zero. When the equivalent encoder is in $\SZ$, then the original encoder is also in its zero state.  Avoiding trivial cases by assuming that the $x^{m-1}$ and $x^0$ coefficients of $p(x)$ are 1, there are $2^m-1$ equivalent encoder states in addition to $\SZ$ that correspond to the all-zero state of the original encoder.  To see this, consider the top diagram in Fig.~\ref{fig:Equivalent_Encoder} in which the equivalent encoder state is shown as the state of the CRC encoder extended with $\nu$ additional  memories.  Note that whatever state the equivalent encoder is in, there is a sequence of $\nu$ bits that will produce $\nu$ zeros at the output of the CRC encoder that will drive the original encoder state to zero.  The specific set of $\nu$ bits is a function of the $m$-bits of state in the CRC encoder.  Thus for any $m$-bits pattern there is a corresponding ($m+\nu$)-bit state of the equivalent encoder that corresponds to the original encoder being in the zero state.

We call the $2^m-1$ non-zero equivalent encoder states for which the original encoder state is zero \emph{detectable-zero states}, forming a set $\SD$, because they correspond to the termination of a detectable error event in the original encoder. The remaining $2^{m+\nu}-2^m$ states of the equivalent code are called \emph{non-zero states}, forming a set $\SN$, because the corresponding states of the original code are not zero.  To terminate an error event in the original encoder, the trellis of the equivalent code transitions from a state in  $\SN$ to $\SZ$ or to a state in $\SD$.  If the transition is to $\SD$ the cumulative errors are detectable because the portion of $q'(x)p(x)$ till this moment is not divisible by $p(x)$.  If the transition is to $\SD$, the cumulative errors are detectable because the portion of $q'(x)p(x)$ till this moment is divisible by $p(x)$.

\begin{table}[!t]
\ifCLASSOPTIONdraftcls
\renewcommand{\arraystretch}{0.6}
\else
\renewcommand{\arraystretch}{1.1}
\fi
\caption{An Example of State Types with $p(x)=x^2+x+1$ and $\nu=2$.}
\label{tbl:State_Type_Example}
\centering
\begin{tabular}{c|c|c|c||c|c}
 & State & \multicolumn{2}{c||}{Equivalent Code} & \multicolumn{2}{c}{Original Code}\\
\cline{3-6}
Time & Type & State & Input & State & Input\\
\hline
&&&&&\\[-2.4ex]
0 & $\SZ$ & \;0000\; & 1 & 00 & 1\\
1 & $\SN$ & 0001 & 1 & 01 & 0\\
2 & $\SN$ & 0011 & 0 & 10 & 0\\
3 & $\SD$ & 0110 & 1 & 00 & 0\\
4 & $\SD$ & 1101 & 0 & 00 & 1\\
5 & $\SN$ & 1010 & 0 & 01 & 1\\
6 & $\SN$ & 0100 & 0 & 11 & 0\\
7 & $\SN$ & 1000 & 0 & 10 & 0\\
8 & $\SZ$ & 0000 &   & 00 &
\end{tabular}
\end{table}
Table~\ref{tbl:State_Type_Example} shows an example using the set-up shown in Fig.~\ref{fig:Equivalent_Encoder} with $p(x)=x^2+x+1$, $q'(x)=x^3+x^2+1$, and $\nu=2$. The original encoder $\boldsymbol{c}(x)$  does not need to be specified for the results in Table~\ref{tbl:State_Type_Example} because any feedforward encoder will produce the same state sequence.  The zero state of the original code is visited at time $0$, $3$, $4$, and $8$. At time $0$, the state begins from $\SZ$. At time $3$, the first $\boldsymbol{c}(x)$ error event ends. This error event is detectable if the codeword ends at time $3$ because the input to the original code, i.e. $x^2$, is not divisible by $p(x)$ . The state remains in $\SD$ at time $4$.  At time 5 a second $\boldsymbol{c}(x)$ error event  begins.  At time $8$, the  $\boldsymbol{c_{eq}}(x)$ state returns to $\SZ$, which is not only the end of the second $\boldsymbol{c}(x)$ error event but also the end of an undetectable double error.  

Note that $\boldsymbol{c_{eq}}(x)$ is catastrophic because its generator polynomials have a common factor $p(x)$.    The catastrophic behavior is expressed through the zero distance loops that occur as the equivalent encoder traverses a sequence of states in $\SD$ while the original convolutional encoder stays in the zero state.  Thus time spent in $\SD$  between $\boldsymbol{c}(x)$ error events can lengthen an undetectable error without increasing its distance.

\subsubsection{Error Events in the Equivalent Encoder}
\label{sec:Error_Events_in_the_Equivalent_Encoder}
Since the all-zero codeword is assumed to be sent, the correct path remains in $\SZ$ forever. An error event in the equivalent encoder occurs if the trellis path leaves $\SZ$ and returns $\SZ$ without any visits to $\SZ$ in between. We will classify these error events according to the number of times it enters $\SD$ from $\SN$ during the deviation. If a trellis path enters $\SD$ from $\SN$ $s-1$ times, then it is classified as an undetectable $s$-tuple error. In an undetectable $s$-tuple error, there are exactly $s$ segments of consecutive transitions between states in $\SN$, which correspond to $s$ error events of the original encoder.  These segments are separated by segments of consecutive transitions between states in $\SD$.

\begin{figure}[!t]
\centering
\ifCLASSOPTIONonecolumn
\includegraphics[width=0.53\columnwidth]{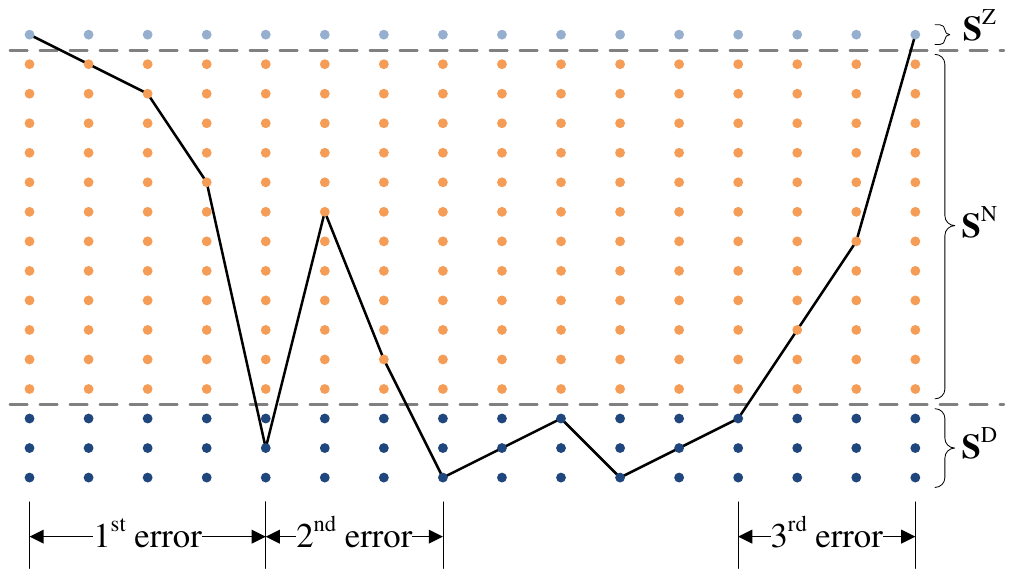}
\else
\includegraphics[width=0.99\columnwidth]{./figure/Event_Classification_small}
\fi
\caption{A trellis diagram of an equivalent convolutional code with $\nu=2$ and $m=2$ having an undetectable triple error. The states are reordered such that the all-zero state is at the top and the detectable-zero states are at the bottom.}
\label{fig:Event_Classification}
\end{figure}
Fig.~\ref{fig:Event_Classification} illustrates an undetectable triple error in a system with $\nu=2$ and $m=2$. The three error events of the original code are separated by visits to $\SD$. The path can leave $\SD$ right after entering it as shown between the first and the second error events; the path can also stay in $\SD$ for a while and then leave $\SD$ from a state different from where it enters $\SD$ as shown between the second and the third error events. Note that $\SZ$ is never directly connected to any state of $\SD$.

The probability of undetectable single errors is bounded in a similar way as \eqref{eqn:PUD_Single_Bound} by
\ifCLASSOPTIONonecolumn
\begin{equation}
\textnormal{P}_{\textnormal{UD}, 1}\leq \textnormal{P}_{>\hat d,1}+\!\sum_{d \in \mathbf{D}_{\hat d, 1}} {\sum_{i = 1}^{a_d^\textnormal{ZZ}} {\max\left\{0, n + \nu - l_{d,i}^\textnormal{ZZ} + 1 \right\}\Prob\left( d \right)}}
\label{eqn:PUD2_Single_Bound},
\end{equation}
\else
\begin{IEEEeqnarray}{l}
\textnormal{P}_{\textnormal{UD}, 1}\!\leq\textnormal{P}_{>\hat d,1}\!+\!\!\sum_{d \in \mathbf{D}_{\hat d, 1}} {\!\sum_{i = 1}^{a_d^\textnormal{ZZ}} {\max\!\left\{\!0, n + \nu - l_{d,i}^\textnormal{ZZ} + 1 \!\right\}\!\Prob\left( d \right)}}
\label{eqn:PUD2_Single_Bound},
\IEEEeqnarraynumspace
\end{IEEEeqnarray}
\fi
where $a_d^\textnormal{ZZ}$ is the number of error events with distance $d$ starting from $\SZ$ and ending at $\SZ$ while never traversing a state in $\SD$.  The length $l_{d,i}^\textnormal{ZZ}$ is the length of the $i^\textnormal{th}$ error event counted in $a_d^\textnormal{ZZ}$. Note that this expression requires the distance spectrum $a_d^\textnormal{ZZ}$ and length spectrum $l_{d,i}^\textnormal{ZZ}$ for $d\in\left[\dfree, \hat d\right]$ obtained using computer search. 

Unlike \eqref{eqn:PUD_Single_Bound}, \eqref{eqn:PUD2_Single_Bound} does not need to check the divisibility of error events. Hence, there is no need to store the error patterns anymore but only their distances and lengths. Consequently the search depth $\hat d$ can go beyond $\tilde d$.

Let $\textnormal{S}^\textnormal{D}_i$ be the $i^\textnormal{th}$ state in $\SD$, where $i\in \left[1,2^m-1\right]$. Define $\mathbf{\Delta}_i$ as the subset of $\SD$ composed of all states connected with $\textnormal{S}^\textnormal{D}_i$ through a path that only includes states in $\SD$. Appendix~\ref{sec:Appendix_Subset_Cyclotomic_Coset} shows that $\mathbf{\Delta}_i$ and the $x$-cyclotomic coset modulo $p(x)$ discussed in Appendix~\ref{sec:Appendix_Search_Double_Error} are equivalent.  Thus if $p(x)$ is a primitive polynomial, all sets $\mathbf{\Delta}_i$ are identical and $\left|\mathbf{\Delta}_i\right|=2^m-1$ is maximized. This may lead to fewer undetectable double errors as shown below and is desirable in the design of a CRC code.

Similar to \eqref{eqn:PUD_Double_Bound2}, the probability of undetectable double errors can be bounded by
\ifCLASSOPTIONonecolumn
\begin{IEEEeqnarray}{C}
\textnormal{P}_{\textnormal{UD}, 2} \leq \textnormal{P}_{>\hat d,2} + \!\!\!\!\!\sum_{\left(d_1 ,d_2\right) \in \mathbf{D}_{\hat d, 2}} {\!\!\!\!\!\!\Prob \left(d_1 + d_2\right)\!\sum_{\phi = 1}^{2^m-1} {\sum_{\textnormal{S}^\textnormal{D}_{\psi} \in \mathbf{\Delta}_{\phi}}{\sum_{i_1=1}^{a_{d_1}^{\textnormal{ZD},\phi}}}{\sum_{i_2=1}^{a_{d_2}^{\textnormal{DZ},\psi}}}}}
\!\sum_{t_1=0}^{\left\lfloor \!\frac{n+\nu-l^{\min}}{\left|\mathbf{\Delta}_{\phi}\right|}\!\right\rfloor}{\!\!\!\!\!\!\left(n+\nu-l^{\min}\!-\left|\mathbf{\Delta}_{\phi}\right|t_1+1\right)}
\label{eqn:PUD2_Double_Bound},\IEEEeqnarraynumspace
\end{IEEEeqnarray}
\else
\begin{IEEEeqnarray}{rCl}
\textnormal{P}_{\textnormal{UD}, 2} &\leq& \textnormal{P}_{>\hat d,2} + \!\!\!\sum_{\left(d_1 ,d_2\right) \in \mathbf{D}_{\hat d, 2}} {\!\!\!\!\!\!\Prob \left(d_1 + d_2\right)\!\sum_{\phi = 1}^{2^m-1} {\sum_{\textnormal{S}^\textnormal{D}_{\psi} \in \mathbf{\Delta}_{\phi}}{\sum_{i_1=1}^{a_{d_1}^{\textnormal{ZD},\phi}}}{\:\sum_{i_2=1}^{a_{d_2}^{\textnormal{DZ},\psi}}}}}\nonumber\\
&&\!\!\sum_{t_1=0}^{\left\lfloor \left(n+\nu-l^{\min}\right)\slash \left|\mathbf{\Delta}_{\phi}\right| \right\rfloor}{\!\!\!\!\!\left(n+\nu-l^{\min}-\left|\mathbf{\Delta}_{\phi}\right|t_1+1\right)}
\label{eqn:PUD2_Double_Bound},
\end{IEEEeqnarray}
\fi
where
\begin{equation}
l^{\min} = l^{\textnormal{ZD},\phi}_{d_1,i_1}+l^{\textnormal{DZ},\psi}_{d_2,i_2}+\delta_{\phi,\psi}
\label{eqn:L_min_Definition}
\end{equation}
is the shortest possible length of the undetectable double error specified by $\left(d_1,d_2,\phi,\psi,i_1,i_2\right)$. 

In \eqref{eqn:PUD2_Double_Bound}, $\phi$ is the index of the state at which the trellis enters $\SD$ at the end of the first error event, and $\psi$ is the index of the state at which the trellis leaves $\SD$ at the beginning of the second error event. In \eqref{eqn:L_min_Definition}, $\delta_{\phi,\psi} < \left|\mathbf{\Delta}_\phi\right|$ is the number of hops required to go from $\textnormal{S}^\textnormal{D}_\phi$ to $\textnormal{S}^\textnormal{D}_\psi$ without leaving $\SD$ for $\textnormal{S}^\textnormal{D}_\psi \in \mathbf{\Delta}_\phi$, where $\delta_{\phi,\psi}=0$ if $\phi=\psi$. The number of error events starting at $\SZ$ and ending at $\textnormal{S}^\textnormal{D}_{\phi}$ with distance $d_1$ is $a^{\textnormal{ZD},\phi}_{d_1}$, and the ${i_1}^\textnormal{th}$ error event of them has length $l^{\textnormal{ZD},\phi}_{d_1,i_1}$, where both numbers are obtained by computer search. The variables $a_{d_2}^{\textnormal{DZ},\psi}$ and $l^{\textnormal{DZ},\psi}_{d_2,i_2}$ are defined in a similar way while the error event starts in $\textnormal{S}^\textnormal{D}_{\psi}$ and ends in $\SZ$. Furthermore, $t_1$ specifies the number of cycles the trellis stays in $\mathbf{\Delta}_{\phi}$ and its upper limit makes sure that the total length of the undetectable double error does not exceed $n+\nu$, the number of trellis stages in the codeword.  As in \eqref{eqn:PUD_Double_Bound2} $\textnormal{P}_{>\hat d,2}$ can often be neglected because terms with $d_1+d_2>\hat{d}$ have negligible probability. 

Although the number and lengths of error events connecting $\SZ$ and states in $\SD$ are obtained by computer searches, the required search depth is only $\hat d - \dfree$. Moreover, $a^{\textnormal{ZD},\phi}_{d_1}$ and $l^{\textnormal{ZD},\phi}_{d_1,i_1}$ for all $\phi$ can be found while searching for $a_d^\textnormal{ZZ}$ and $l_{d,i}^\textnormal{ZZ}$ because these two types of error events both start from $\SZ$. Regarding $a_{d_2}^{\textnormal{DZ},\psi}$ and $l^{\textnormal{DZ},\psi}_{d_2,i_2}$, the associated error events start from $2^m-1$ different states and can be found through $2^m-1$ separated searches. Nevertheless, since these error events end at $\SZ$, only one backward search is necessary to capture all of them. In the backward search, bits in shift registers move backward. One can simply treat $x$ as $x^{-1}$ in polynomial representations and apply the same search algorithm. 
%An easier way to search for error events ending at $\SZ$ is to utilize the search result of error events starting from $\SZ$. See Appendix~\ref{sec:Appendix_Smart_Search_Using_Delta} for detail.

\subsubsection{Undetected Error Probability}
\label{sec:Inclusive_Undetected_Error_Probability}
As shown in Fig.~\ref{fig:Event_Classification}, an undetectable $s$-tuple error is composed of several parts: one error event from $\SZ$ to a state in $\SD$, $s-2$ error events from a state in $\SD$ to a state in $\SD$ with visits to $\SN$ in between, the final error event from a state in $\SD$ to $\SZ$, and also $s-1$ paths inside $\SD$ connecting consecutive error events. Note that in the example of Fig.~\ref{fig:Event_Classification}, where $s=3$, where the path connecting the first and the second error events has a length of zero. Let $\phi_u$ and $\psi_u$ be the indices of the start and end states of the $u^\textnormal{th}$ transitions in $\SD$, respectively. In Fig.~\ref{fig:Event_Classification}, we have $\phi_1=\psi_1=2$ for the first transition; the second transition starts from $\textnormal{S}^\textnormal{D}_{\phi_2}$ and ends at $\textnormal{S}^\textnormal{D}_{\psi_2}$, where $\phi_2=3$ and $\psi_2=1$. Also, let the number of error events started at $\textnormal{S}^\textnormal{D}_{\psi_{u-1}}$ and ended at $\textnormal{S}^\textnormal{D}_{\phi_u}$ with distance $d_u$ be $a^{\textnormal{DD},\psi_{u-1},\phi_u}_{d_u}$ and the length of the ${i_u}^\textnormal{th}$ error event of them be $l^{\textnormal{DD},\psi_{u-1},\phi_u}_{d_u,i_u}$, where both numbers are obtained by computer search. Although we need to perform $2^m-1$ separated searches to obtain all $a^{\textnormal{DD},\psi_{u-1},\phi_u}_{d_u}$ and $l^{\textnormal{DD},\psi_{u-1},\phi_u}_{d_u,i_u}$, a search depth of $\hat d - \left(s-1\right)\dfree$ is sufficient. 
%Similar to the search for error events ending at $\SZ$, there is an easier approach presented in Appendix~\ref{sec:Appendix_Smart_Search_Using_Delta}. 

To simplify the notation, define the following sets:
\ifCLASSOPTIONonecolumn
\begin{IEEEeqnarray*}{rCl}
\mathbf{\Phi}_s &=& \left\{\left(\phi_1,\cdots,\phi_s \right)\left| \phi_u \in \left[1,2^m-1\right]\:\:\forall u\in\left[1,s\right]\right.\right\}\\
\mathbf{\Psi}_s &=& \left\{\left(\psi_1,\cdots,\psi_s \right)\left| \psi_u \in \mathbf{\Delta}_{\phi_u}\:\:\forall u\in\left[1,s\right]\right.\right\}\\
\mathbf{I}'_s &=&\left\{\left(i_1,\cdots,i_s \right)\left| i_1\in\left[1,a^{\textnormal{ZD},\phi_1}_{d_1}\right]\!, i_s\in\left[1,a^{\textnormal{DZ},\psi_{s-1}}_{d_s}\right]\!, i_u \in \left[1,a^{\textnormal{DD},\psi_{u-1},\phi_u}_{d_u}\right]\:\forall u\in\left[2,s-1\right]\right.\right\}\\
\mathbf{T}_s &=& \left\{\!\left(t_1,\cdots,t_s \right)\left|t_u \geq 0\:\;\forall u\in\left[1,s\right],\sum_{u=1}^s {\left|\mathbf{\Delta}_{\phi_u}\right|t_u}\leq n+r-l^{\min}_{s+1}\right.\right\}\!,
\end{IEEEeqnarray*}
\else
\begin{IEEEeqnarray*}{rCl}
\mathbf{\Phi}_s &=& \left\{\left(\phi_1,\cdots,\phi_s \right)\left| \phi_u \in \left[1,2^m-1\right]\:\:\forall u\in\left[1,s\right]\right.\right\}\\
\mathbf{\Psi}_s &=& \left\{\left(\psi_1,\cdots,\psi_s \right)\left| \psi_u \in \mathbf{\Delta}_{\phi_u}\:\:\forall u\in\left[1,s\right]\right.\right\}\\
\mathbf{I}'_s &=&\left\{\left(i_1,\cdots,i_s \right)\left| i_1\in\left[1,a^{\textnormal{ZD},\phi_1}_{d_1}\right]\!, i_s\in\left[1,a^{\textnormal{DZ},\psi_{s-1}}_{d_s}\right]\right.\right.\!\!\!,\\
\IEEEeqnarraymulticol{3}{r}{\left.i_u \in \left[1,a^{\textnormal{DD},\psi_{u-1},\phi_u}_{d_u}\right]\:\:\forall u\in\left[2,s-1\right]\right\}}\\
\mathbf{T}_s &=& \Bigg\{\!\left(t_1,\cdots,t_s \right)\Bigg|t_u \geq 0\;\;\;\forall u\in\left[1,s\right],\\
\IEEEeqnarraymulticol{3}{r}{\sum_{u=1}^s {\left|\mathbf{\Delta}_{\phi_u}\right|t_u}\leq n+\nu-l^{\min}_{s+1}\Bigg\},\!\!}
\end{IEEEeqnarray*}
\fi
where $l^{\min}_s$ is the shortest possible length of the undetectable $s$-tuple error specified in a similar way as \eqref{eqn:L_min_Definition} and given by
\begin{IEEEeqnarray}{c}
l^{\min}_s=l^{\textnormal{ZD},\phi_1}_{d_1,i_1}\!+\!\sum_{u=2}^{s-1}{l^{\textnormal{DD},\psi_{u-1},\phi_u}_{d_u,i_u}}+l^{\textnormal{DZ},\psi_{s-1}}_{d_s,i_s}+\!\sum_{u=1}^{s-1}{\delta_{\phi_u,\psi_u}}
\label{eqn:L_min_s_Definition}.
\ifCLASSOPTIONtwocolumn
\IEEEeqnarraynumspace
\fi
\end{IEEEeqnarray}

Similar to \eqref{eqn:PUD_General_Bound}, the probability of having an undetectable error is now bounded by
\ifCLASSOPTIONonecolumn
\begin{IEEEeqnarray}{rCl}
\textnormal{P}_\textnormal{UD}\!&\leq\!&\sum_{s = 2}^\infty {\,\sum_{\left(d_1, \cdots ,d_s \right) \in \mathbf{D}_{\hat d, s}} {\Prob\left(\sum_{u = 1}^s {d_u}\right)\sum_{\left(\phi_1,\cdots,\phi_{s-1}\right)\in\mathbf{\Phi}_{s-1}}{\;\sum_{\left(\psi_1,\cdots,\psi_{s-1}\right)\in\mathbf{\Psi}_{s-1}}{\;\sum_{\left(i_1, \cdots ,i_s \right) \in \mathbf{I}'_s } {\;\sum_{\left(t_1, \cdots ,t_{s-1} \right) \in \mathbf{T}_{s-1}}}}}}}\nonumber\\
&&\!\!\!\left(\!n\!+\!\nu\!-\!l^{\min}_s\!-\!\sum_{u=1}^{s-1}{\left|\mathbf{\Delta}_{\phi_u}\right|t_u}\!+\!1\!\right)\!\!+\!\!\!\sum_{d_1 \in \mathbf{D}_{\hat d, 1}} {\sum_{i = 1}^{a_{d_1}^\textnormal{ZZ}} {\max\!\left\{0, n\!+\!\nu\!-\!l_{d_1,i}^\textnormal{ZZ}\!+\!1\right\}\!\Prob\left( d_1 \right)}}\!+\!\sum_{s=1}^\infty{\textnormal{P}_{>\hat d,s}}
\label{eqn:PUD2_General_Bound}.
\end{IEEEeqnarray}
\else
\begin{IEEEeqnarray}{rCl}
\textnormal{P}_\textnormal{UD} &\leq&\sum_{d_1 \in \mathbf{D}_{\hat d, 1}} {\sum_{i = 1}^{a_{d_1}^\textnormal{ZZ}} {\max\!\left\{0, n + \nu - l_{d_1,i}^\textnormal{ZZ}+ 1 \right\}\Prob\left( d_1 \right)}}\nonumber\\
&&+\sum_{s = 2}^\infty {\,\sum_{\left(d_1, \cdots ,d_s \right) \in \mathbf{D}_{\hat d, s}} {\!\Prob\left(\sum_{u = 1}^s {d_u}\right)\!\sum_{\left(\phi_1,\cdots,\phi_{s-1}\right)\in\mathbf{\Phi}_{s-1}}}}\nonumber\\
&&\sum_{\left(\psi_1,\cdots,\psi_{s-1}\right)\in\mathbf{\Psi}_{s-1}}{\;\sum_{\left(i_1, \cdots ,i_s \right) \in \mathbf{I}'_s } {\;\sum_{\left(t_1, \cdots ,t_{s-1} \right) \in \mathbf{T}_{s-1}}}}\nonumber\\
&&\left(n+\nu-l^{\min}_s-\sum_{u=1}^{s-1}{\left|\mathbf{\Delta}_{\phi_u}\right|t_u}+1\right)\!+\!\sum_{s=1}^\infty{\textnormal{P}_{>\hat d,s}}
\label{eqn:PUD2_General_Bound}.\IEEEeqnarraynumspace
\end{IEEEeqnarray}
\fi
The last term is the probability sum of all large-distance errors, which are assumed to be undetectable, and can be calculated using \eqref{eqn:P_tilde_Sum_Bound_3}. By letting $\textnormal{P}_{>\hat d,s}=0$, we obtain an approximation. By letting all $l^\textnormal{ZZ}$, $l^\textnormal{ZD}$, $l^\textnormal{DD}$, and $l^\textnormal{DZ}$ equal to $\nu+1$, we obtain a looser bound which does not require any length information. These two techniques are applicable to every $\textnormal{P}_{\textnormal{UD}, s}$, including \eqref{eqn:PUD2_Single_Bound} and \eqref{eqn:PUD2_Double_Bound}.

\subsection{Comparison}
\label{sec:Compare_Both_Methods}
The main benefit of the construction method is that it is often able to search deeper than the exclusion method because the output pattern is not required.  However, the required memory size scales with the number of states $2^{m+\nu}$ rather than $2^{\nu}$ so this approach can encounter difficulty in analyzing high-order CRC codes. In contrast, the error events searched in the exclusion method belong to the original convolutional code, whose number of states is just $2^\nu$ and is independent of the degree of the CRC code. In fact, both methods create the dominant parts (till $\tilde{d}$ or $\hat{d}$) of the distance spectrum of the equivalent catastrophic convolutional code with finite length. As explained in Section~\ref{sec:Search_for_CRC}, we found it useful to draw on both approaches as we searched for optimal CRC polynomials for a specific convolutional code. The exclusion method is suitable for short-distance ($d\leq\tilde{d}$) undetectable single errors, and the construction method is preferred when we search for longer undetectable single errors and double errors.

\begin{figure}[!t]
\centering
\ifCLASSOPTIONonecolumn
\includegraphics[width=0.53\columnwidth]{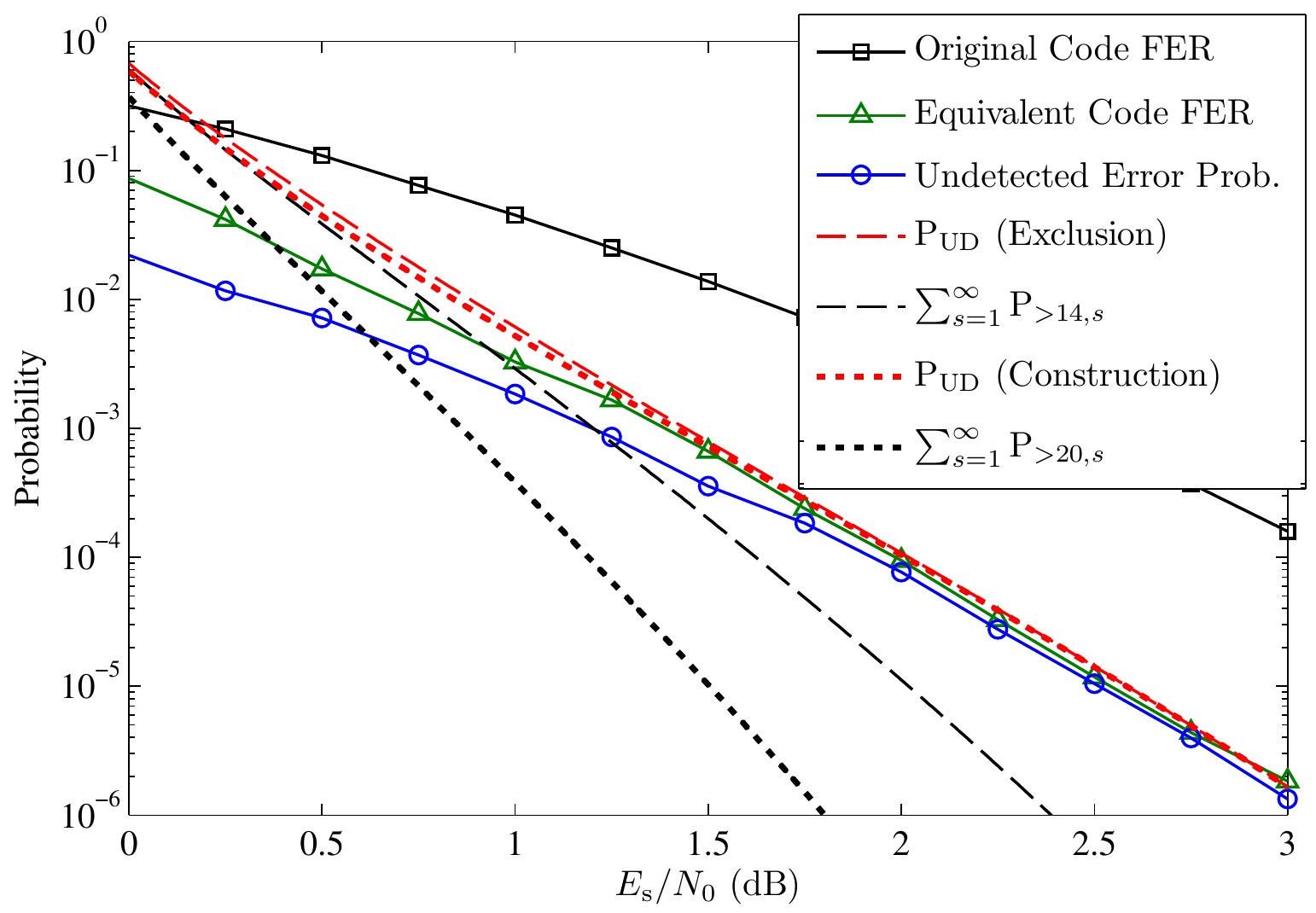}
\else
\includegraphics[width=0.99\columnwidth]{./figure/Catastrophic_FER}
\fi
\caption{A comparison of the simulated undetected error probability with the simulated frame error rate of the equivalent code and the analyses from the exclusion and construction methods. The system is equipped with the CRC code $p(x)=x^3+x+1$ and the convolutional code $(23,35)_8$. The CRC codeword length is $n=1024$ bits.}
\label{fig:Catastrophic_FER_vs_SNR}
\end{figure}

Fig.~\ref{fig:Catastrophic_FER_vs_SNR} compares the simulated undetected error probability to the bounds produced by the exclusion and construction methods.  We consider the CRC code $p(x)=x^3+x+1$ concatenated with the memory size $\nu=4$ convolutional code with generator polynomial $(23,35)_8$ in octal and $\dfree=7$. The information length is $k=1021$ bits and thus the CRC codeword length is $n=1024$ bits. The FER of this original convolutional code is plotted as a reference. The equivalent catastrophic convolutional code is $(255, 317)_8$ and its FER is also simulated. 

We can see from Fig.~\ref{fig:Catastrophic_FER_vs_SNR} that the undetected error probability is upper bounded by the FER of the equivalent code, which equals $\textnormal{P}_\textnormal{UD}$, the probability of having an undetectable error in the CRC codeword, and this bound gets tighter as SNR increases.  The equivalent code FER is above the probability of undetected error because it is possible that a frame has \emph{both} an undetectable error event and a detectable error event, which causes a frame error in the equivalent code but does not cause an undetected error in the concatenated CRC and convolutional codes. 

The upper bounds of $\textnormal{P}_\textnormal{UD}$ are computed using the exclusion method \eqref{eqn:PUD_General_Bound} and construction method \eqref{eqn:PUD2_General_Bound}. In our calculation, the search depth limit of the exclusion method is $\tilde d = 14$, and $\hat d = 20$ is the depth limit for the construction method.  Since $\dfree=7$, only undetectable single and double errors are considered. The probability sum of all large-distance terms given by \eqref{eqn:P_tilde_Sum_Bound_3} are plotted to verify that they are negligible, except for the exclusion method at SNR below $0.75$ dB. Although the construction  method used a deeper search, it is still quite close to the exclusion method even in the low SNR region. It can be seen that these analysis methods deliver accurate bounds at high SNR for both the undetected error probability and the FER of the catastrophic code.

\section{Search Procedure for Optimal CRC Codes}
\label{sec:Search_for_CRC}
In this section, we will present an efficient way to find the optimal degree-$m$ CRC code for a targeted convolutional code and information length $k$. Note that the performance of a CRC code depends on the information length \cite{Koopman_CRC_Length}.  A CRC code may be powerful for short sequences but have numerous undetectable long errors that are produced by a specific convolutional code.

Since the coefficients of $x^m$ and $x^0$ terms are both one, there are $2^{m-1}$ candidates of degree-$m$ CRC generator polynomials $p(x)$.  In principle, either the exclusion method or the construction method can produce the undetected error probability for each candidate allowing selection of the best $p(x)$. However, this process is time-consuming if $m$ is large. Both exclusion and construction methods need to compute the distance spectrum of undetectable errors up to some distance $d$. We can reduce the computation time by skipping the distance spectrum searches of suboptimal CRC codes.

When the FER is low, the undetectable error rate of a CRC code is dominated by the undetectable errors with the smallest distance. Let the smallest distance of the undetectable errors be $d_{\min}$. We can evaluate a CRC-convolutional concatenated code by its distance spectrum at around $d_{\min}$. To be more precise, a polynomial should be removed from the candidate list if it has a smaller $d_{\min}$ than the others or if it has more undetectable errors associated to the same $d_{\min}$. In a convolutional code, since the number of error events $a_d$ grows exponentially as the associated distance $d$ increases, the cost to find all undetectable errors grows exponentially as well. Hence, the CRC code search starts with $d=\dfree$ and updates the candidate list by keeping only the CRC generator polynomials with the fewest undetectable errors. Next, repeat the procedure with the next higher $d$ until only one polynomial remains in the list.

When $d < 2\dfree$, only single errors are possible.  The exclusion method can count the number of undetectable single errors of each candidate when $d\leq \tilde d$.  We can perform a computer search for error events of the original code and check the divisibility of each of them. Note that if an error event is found to be undetectable, all of its possible offsets in the codeword should be counted. Since all candidates check the same set of error events, only one computer search with multiple divisibility checks (one for each CRC code) is sufficient. In contrast, using the construction method requires construction of equivalent encoders for each candidate separately. Hence, for the initial values of $d$ near $\dfree$, checking the divisibility via the exclusion method is preferred. Of course, once $d > \tilde d$, searching for undetectable single errors of the equivalent codes as in the construction method is the only approach.

When $d \geq 2\dfree$, undetectable double errors need to be considered in addition to single errors. The divisibility test should be applied to all combinations of error event patterns $e_1(x)$, $e_2(x)$, and their gaps $g_1$. Even if the concept of cyclotomic cosets discussed in Appendix~\ref{sec:Appendix_Search_Double_Error} is utilized, we still need to construct all cyclotomic cosets through about $2^m$ divisions and also check if each of the remainder of $e_2(x)$ divided by $p(x)$ is in the same cyclotomic coset as the remainder of $e_1(x)$ divided by $p(x)$. 

Alternatively, undetectable double errors can be directly created using the construction method. In the construction method, the error events connecting $\SZ$ and states in $\SD$ with distances between $\dfree$ and $d-\dfree$ are found through computer searches. In fact, these events can be generated using the detectable error events of the original code previously found by exclusion if $d-\dfree\leq\tilde d$. For example, since the detectable error pattern $e_1(x)$ is known, the corresponding error event in the equivalent encoder trellis starts from $\SZ$ and traverses the trellis with the input sequence given by the quotient of $x^me_1(x)$ divided by $p(x)$. The state $\textnormal{S}^\textnormal{D}_{\phi}$, where it ends, is thus determined by the last $m+\nu$ input bits, and its length $l^{\textnormal{ZD},\phi}_{d_1,i_1}$ has already been provided by the degree of $e_1(x)$. 

For error events starting from states in $\SD$ and ending in $\SZ$ with pattern $e_2(x)$ previously obtained by exclusion, traverse the trellis in reverse from $\SZ$ with the input sequence given by the quotient of $x^{m+\nu+l'}e_2(x^{-1})$ divided by $x^mp(x^{-1})$, where $l'=l^{\textnormal{DZ},\psi}_{d_2,i_2}-1$ is the degree of $e_2(x)$. Note that $x^{l'}e_2(x^{-1})$ and $x^mp(x^{-1})$ are the reverse bit-order polynomial representations of $e_2(x)$ and $p(x)$, respectively. The state $\textnormal{S}^\textnormal{D}_{\psi}$ where the error event begins is determined by the last $m+\nu$ input bits in reverse order.  

%An easier way to determine $\textnormal{S}^\textnormal{D}_{\psi}$, requiring no additional polynomial divisions, is presented in Appendix~\ref{sec:Appendix_Smart_Search_Using_Delta}. Thus, to count the number of undetectable double errors, creating them directly as in the construction method is preferred.

According to the discussion at the end of Appendix~\ref{sec:Appendix_Search_Double_Error}, $d_{\min}$ is not likely to be much greater than $2\dfree$ when information length $k$ is long enough. In other words, the CRC code search algorithm is usually finished before reaching $3\dfree$ and does not need to count the number of undetectable triple errors.

\section{CRC Design Example for $\nu=6$, $k=1024$}
\label{sec:Numerical_Result}
As an example we present the best CRC codes of degree $m\leq 16$ specifically for the popular memory size $\nu=6$ convolutional code with generator polynomial $(133,171)_8$ with information length $k=1024$ bits. Note that the proposed design method is applicable to all convolutional codes and information lengths, and not limited to the choices used for this example. The corresponding undetected error probability is also calculated and compared with existing CRC codes.

\begin{table}[!t]
\ifCLASSOPTIONdraftcls
\renewcommand{\arraystretch}{0.6}
\else
\renewcommand{\arraystretch}{1.1}
\fi
\caption{Standard CRC Codes or CRC Codes Recommended by Koopman and Chakravarty (K\&C) \cite{Koopman_CRC_Length}, and the Best CRC Codes for Convolutional Code $(133,171)_8$ with $k=1024$ Bits.}
\label{tbl:CRC_List}
\centering
\begin{tabular}{c|lc|r@{\;\;}r@{\;\;}r@{\;\;}r@{\;\;}r@{\;\;}r@{\;\;}r}
 & Gen. & \multicolumn{8}{|r}{Undetectable Single Distance Spectrum $a_d^\textnormal{ZZ}$}\\[0.5pt]
\cline{3-10}
&&\multicolumn{1}{|c|}{}&&&&&&&\\[-2.3ex]
Name & Poly. & \multicolumn{1}{|c|}{$d$} & 10 & 12 & 14 & 16 & 18 & 20 & 22\\
\hline
\hline
&&&&&&&&&\\[-2.6ex]
K\&C-3  & 0x5    & & 1 & 5 & 19 & 170 & 941 & 5050 & 29290\\
Best-3     & 0x7    & & 0 & 7 & 24 & 169 & 879 & 5111 & 29363\\
\hline
&&&&&&&&&\\[-2.6ex]
CRC-4      & 0xF    & & 1 & 2 & 11 &  79 & 464 & 2504 & 14719\\
Best-4     & 0xD    & & 0 & 1 & 17 &  91 & 462 & 2537 & 14674\\
\hline
&&&&&&&&&\\[-2.6ex]
CRC-5      & 0x15   & & 1 & 2 &  9 &  52 & 267 & 1378 &  8005\\
Best-5     & 0x11   & & 0 & 0 &  4 &  52 & 230 & 1257 &  7275\\
\hline
&&&&&&&&&\\[-2.6ex]
CRC-6      & 0x21   & & 0 & 1 &  4 &  21 & 124 &  572 &  3659\\
Best-6     & 0x29   & & 0 & 0 &  1 &  22 & 124 &  641 &  3650\\
\hline
&&&&&&&&&\\[-2.6ex]
CRC-7      & 0x48   & & 0 & 0 &  1 &  14 &  55 &  298 &  1877\\
Best-7     & 0x47   & & 0 & 0 &  0 &   7 &  70 &  322 &  1867\\
\hline
&&&&&&&&&\\[-2.6ex]
CRC-8      & 0xEA   & & 0 & 0 &  0 &   4 &  36 &  174 &   871\\
Best-8     & 0x89   & & 0 & 0 &  0 &   1 &  29 &  177 &   938\\
\hline
&&&&&&&&&\\[-2.6ex]
K\&C-9  & 0x167  & & 0 & 0 &  0 &   4 &  13 &   73 &   477\\
Best-9     & 0x177  & & 0 & 0 &  0 &   0 &  14 &  104 &   437\\
\hline
&&&&&&&&&\\[-2.6ex]
CRC-10     & 0x319  & & 0 & 0 &  0 &   1 &   8 &   41 &   239\\
Best-10    & 0x314  & & 0 & 0 &  0 &   0 &   3 &   49 &   223\\
\hline
&&&&&&&&&\\[-2.6ex]
CRC-11     & 0x5C2  & & 0 & 0 &  0 &   0 &   7 &   17 &   107\\
Best-11    & 0x507  & & 0 & 0 &  0 &   0 &   0 &   24 &   113\\
\hline
&&&&&&&&&\\[-2.6ex]
CRC-12     & 0xC07  & & 0 & 0 &  0 &   0 &   3 &   12 &    48\\
Best-12    & 0xA10  & & 0 & 0 &  0 &   0 &   0 &    4 &    66\\
\hline
&&&&&&&&&\\[-2.6ex]
K\&C-13 & 0x102A & & 0 & 0 &  0 &   0 &   1 &    7 &    36\\
Best-13    & 0x1E0F & & 0 & 0 &  0 &   0 &   0 &    1 &    29\\
\hline
&&&&&&&&&\\[-2.6ex]
K\&C-14 & 0x21E8 & & 0 & 0 &  0 &   0 &   1 &    2 &    15\\
Best-14    & 0x314E & & 0 & 0 &  0 &   0 &   0 &    0 &    11\\
\hline
&&&&&&&&&\\[-2.6ex]
K\&C-15 & 0x4976 & & 0 & 0 &  0 &   0 &   1 &    1 &     6\\
Best-15    & 0x604C & & 0 & 0 &  0 &   0 &   0 &    0 &     3\\
\hline
&&&&&&&&&\\[-2.6ex]
CRC-16     & 0xA001 & & 0 & 0 &  0 &   0 &   0 &    1 &     3\\
Best-16    & 0x8E61 & & 0 & 0 &  0 &   0 &   0 &    0 &     1\\
\hline
\multicolumn{3}{@{\,}c@{\,}|}{Original Distance Spectrum $a_d$} & 11 & 38 & 193 & 1331 & 7275 & 40406 & 234969
\end{tabular}
\end{table}
Table~\ref{tbl:CRC_List} shows the standard CRC codes listed in \cite{Koopman_CRC_Length} and the best CRC codes found by the search procedure in Section \ref{sec:Search_for_CRC}.   For degrees with no standard codes, those recommended by Koopman and Chakravarty in \cite{Koopman_CRC_Length} are listed and called K\&C. The notation of generator polynomials is in hexadecimal as used in \cite{Koopman_CRC_Length}. For example, CRC-8 has generator polynomial $x^8+x^7+x^6+x^4+x^2+1$ expressed as 0xEA, where the most and least significant bits represent the coefficients of $x^8$ and $x^1$ terms, respectively. The coefficient of $x^0$ term is always one and thus omitted. 

Table~\ref{tbl:CRC_List} also gives the distance spectrum of undetectable single errors $a_d^\textnormal{ZZ}$ of each CRC code up to $d=22$. The distance spectrum of the original convolutional code $a_d$ is given as a reference. Note that, since this convolutional code has $\dfree=10$, a smaller $a_{20}^\textnormal{ZZ}$ or $a_{22}^\textnormal{ZZ}$ does not mean fewer undetectable errors at distance $d=20$ or $22$. Undetectable double errors should also be counted for $d\geq 20$ to judge a candidate.

During the search for the best CRC codes with degrees $m\leq 11$, only single errors need to be considered because one candidate will outperform all the others before looking at $d=20$. Although the best degree-$11$ CRC code has $d_{\min}=20$, all the other candidates have $d_{\min}<20$ and are eliminated before the end of the $d=18$ round. Since the lengths of single errors $l_{d,i}$ for $d<20$, ranging from $7$ to $43$, are much shorter than $n+\nu$, a candidate that has fewer types of dominant undetectable error events will have fewer dominant undetectable errors in total. In other words, when undetectable single errors dominate and information length $k$ is long enough, the best CRC code should possess the smallest $a_d^\textnormal{ZZ}$.

When $d_{\min}\geq 2\dfree$, the dominant undetectable errors include double errors. In this case, a smaller $a_d^\textnormal{ZZ}$ does not mean a better code because it only considers single errors. For example, the degree-$16$ polynomials 0xF8F1 and 0x8E61 both have $d_{\min}=22$. The former has $a_{22}^\textnormal{ZZ}=0$ while the latter has $a_{22}^\textnormal{ZZ}=1$. However, at $d=22$, the former has so many ($2860$) undetectable double errors that the number is greater than the total count of undetectable single and double errors ($1011 + 1424$) of the latter, when the information length is $k=1024$ bits. However, when $k=512$ bits, the former has fewer undetectable errors and becomes optimal.

Therefore, different information lengths may lead to different optimal CRC designs. The authors of \cite{Koopman_CRC_Length} have included the information length $k$ as a key design parameter and proposed a methodology to determine ``good'' CRC codes for a range of $k$. The same rule can be applied here. First, find the CRC polynomials possessing the largest $d_{\min}$ for the longest $k$. Then, consider shorter $k$ and keep the CRC polynomials having the largest $d_{\min}$, which might increase as $k$ decreases.
\begin{table}[!t]
\ifCLASSOPTIONdraftcls
\renewcommand{\arraystretch}{0.6}
\else
\renewcommand{\arraystretch}{1.1}
\fi
\caption{The Best CRC Codes for Convolutional Code $(133,171)_8$ with $k=256$, $512$,  or $1024$ Bits. Bold Numbers Indicate the $k$ for which the Code is Designed. The ``Good'' Codes for This Range of $k$ are Specified.}
\label{tbl:CRC_Length}
\centering
\begin{tabular}{c|lc|r@{\,}r@{\;\;\;}r@{\,}r@{\;\;\;}r@{\,}r}
 & Gen. & \multicolumn{7}{|r}{$(d_{\min},\textnormal{ count of undetectable errors at }d_{\min})$}\\
\cline{3-9}
&&\multicolumn{1}{|c|}{}&&&&&&\\[-2.3ex]
Degree & Poly. & \multicolumn{1}{|c|}{$k$} & \multicolumn{2}{c}{256 bits} & \multicolumn{2}{c}{\!\!\!\!512 bits} & \multicolumn{2}{c}{\!\!\!\!1024 bits} \\
\hline
\hline
&&&&&&&&\\[-2.6ex]
\multirow{2}{*}{12} & \multicolumn{2}{|l|}{0xA10 (good)} & \bf(20,& \bf1664) & (20,& 5525) & \bf(20,& \bf17732)\\
                    & \multicolumn{2}{|l|}{0x8DC (good)} & (20,& 1904) & \bf(20,& \bf4748) & (20,& 19283)\\
\hline
&&&&&&&&\\[-2.6ex]
\multirow{2}{*}{13} & \multicolumn{2}{|l|}{0x18F6 (good)} & \bf(20,& \bf169) & (20,& 1474) & (20,& 7452)\\
                    & \multicolumn{2}{|l|}{0x1E0F (good)} & (20,& 289) & \bf(20,& \bf1187) & \bf(20,& \bf5301)\\
\hline
&&&&&&&&\\[-2.6ex]
\multirow{2}{*}{14} & \multicolumn{2}{|l|}{0x2E20}        & \bf(22,& \bf3196) & (20,&   520) & (20,& 2056)\\
                    & \multicolumn{2}{|l|}{0x314E (good)} & (22,& 4698) & \bf(22,& \bf12324) & \bf(20,&  \bf198)\\
\hline
&&&&&&&&\\[-2.6ex]
   & \multicolumn{2}{|l|}{0x6D80}        & \bf(22,& \bf962) & (20,&  253) & (20,&  765)\\
15 & \multicolumn{2}{|l|}{0x76AD}        & (22,& 1210) & \bf(22,& \bf2808) & (20,& 1382)\\
   & \multicolumn{2}{|l|}{0x604C (good)} & (22,& 1767) & (22,& 4414) & \bf(22,& \bf13329)\\
\hline
&&&&&&&&\\[-2.6ex]
   & \multicolumn{2}{|l|}{0xA219}        & \bf(24,& \bf7396) & (22,& 316) & (20,& 454)\\
16 & \multicolumn{2}{|l|}{0xF8F1 (good)} & (24,& 9823) & \bf(22,& \bf219) & (22,& 2860)\\
   & \multicolumn{2}{|l|}{0x8E61}        & (22,&  243) & (22,& 629) & \bf(22,& \bf2435)\\
\end{tabular}
\end{table}
Table~\ref{tbl:CRC_Length} shows the best CRC codes for $k=256$, $512$, or $1024$ bits and identifies the ``good'' codes for this range of information lengths. The bold numbers indicate the $k$ for which the code is designed. For example, the code 0xA10 is the best degree-$12$ code at lengths $k=256$ and $1024$ bits. Note that the codes with degrees $m\leq11$ are not shown since undetectable single errors dominate and thus the best CRC codes for these $k$ are identical. In fact, the best CRC code for the largest $k$ is usually ``good'' for shorter $k$. In our case, the best $k=1024$ CRC codes have no more than $0.1$dB loss compared to the best $k=256$ and the best $k=512$ CRC codes at information lengths $k=256$ and $512$ bits, respectively, except for the degree-$16$ codes.

% Furthermore, if two candidates have the same $a_{d_{\min}}^\textnormal{ZZ}$, the information length $k$ can impact the choice of the best CRC code. For example, the degree-$16$ polynomials 0x90DB and 0x8E61 both have $d_{\min}=22$ and $a_{22}^\textnormal{ZZ}=1$.  0x90DB has one length-$44$ dominant undetectable single error, but the longest error of 0x8E61 is length $36$.  Thus 0x90DB has eight less undetectable single errors at $d=22$ than the polynomial 0x8E61, 0x90DB is not better at this specific information length ($k=1024$) because it has more dominant undetectable double errors: at $d=22$, 0x90DB has $1505$ undetectable double errors containing eight combinations of $e_1(x)$ and $e_2(x)$ with combined lengths ranging from $694$ to $1023$ while  0x8E61 has $1424$ undetectable double errors comprising only three combinations with relatively shorter lengths ranging from $405$ to $754$. Of course, other longer undetectable double errors, whose lengths are greater than $n+\nu$, are not counted. However, when the information length is shorter than $1010$ bits, the polynomial 0x90DB has smaller total count of undetectable errors at $d=22$ and is preferred.

\begin{figure}[!t]
\centering
\ifCLASSOPTIONonecolumn
\includegraphics[width=0.53\columnwidth]{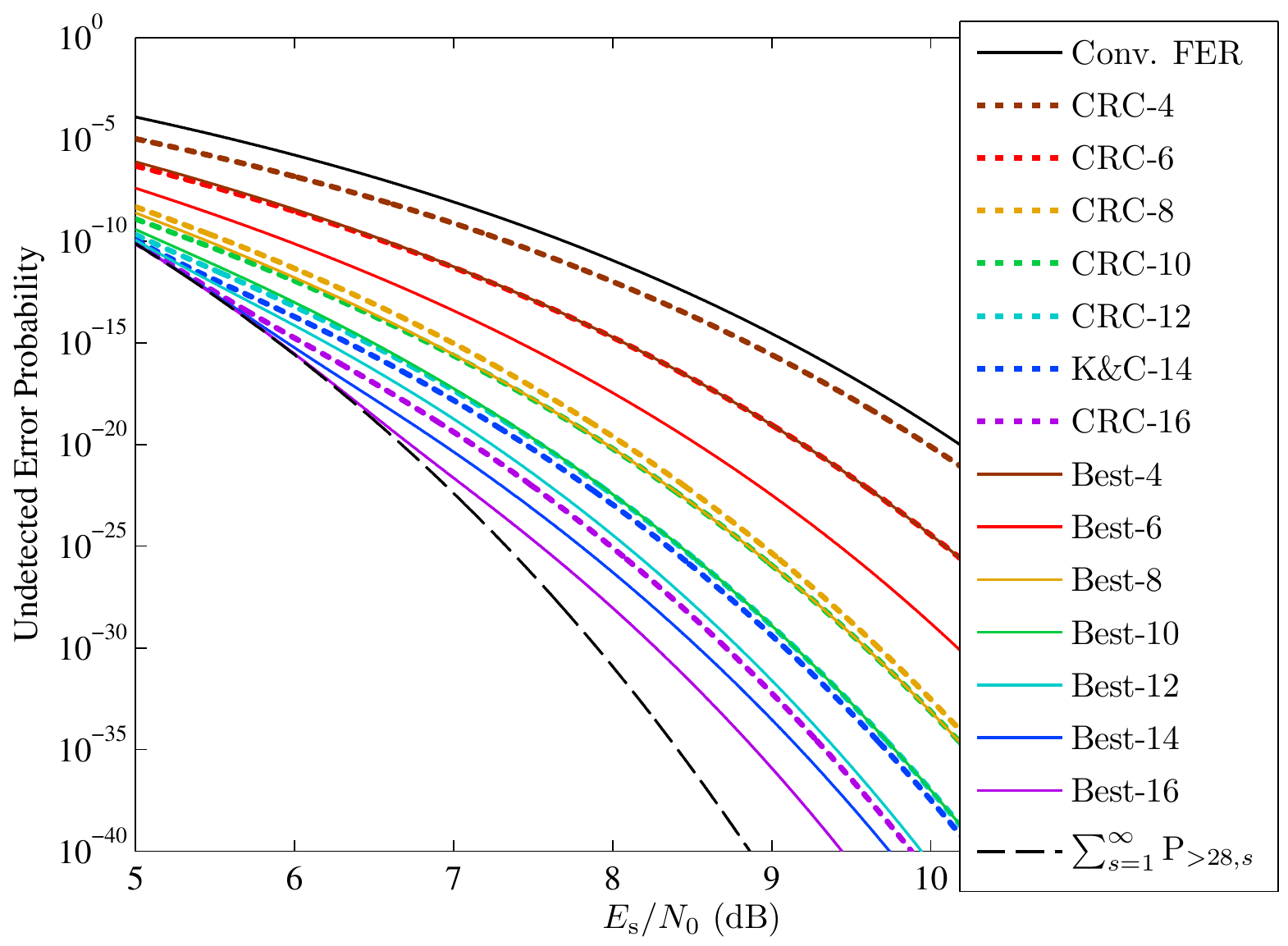}
\else
\includegraphics[width=0.99\columnwidth]{./figure/Undetected_FER}
\fi
\caption{The undetected error probability of the existing and best CRC codes for convolutional code $(133,171)_8$ with information length $k=1024$ bits computed using the construction method.}
\label{fig:Undetected_FER_vs_SNR}
\end{figure}
In Fig.~\ref{fig:Undetected_FER_vs_SNR}, the bounds of undetected error probability of the existing and best CRC codes for information length $k=1024$ bits are shown. For clarity, only even degrees of the CRC codes are displayed. The upper bound of the original convolutional code FER without any CRC code, calculated using transfer function techniques \cite{Convolutional_Code_PER_Milcom}, is plotted as a reference. In our calculation, the search depth limit of the exclusion method is $\tilde d=22$ and not enough for high degree CRC codes. Therefore, the construction method \eqref{eqn:PUD2_General_Bound} was used with $\hat d=28$. 

The probability sum of all large-distance terms calculated using \eqref{eqn:P_tilde_Sum_Bound_3} is also plotted to illustrate that the large-distance terms really are negligible even assuming they are all undetectable, except for the best degree-$16$ CRC code at SNR below $7$ dB and some other codes at SNR below $6$ dB. Note that since the operation $e^{\bar{\textnormal{P}}}-1$ in \eqref{eqn:P_tilde_Sum_Bound_3} causes non-negligible rounding errors in the high SNR region, it was approximated by the first ten terms, which results a similar expression as \eqref{eqn:P_tilde_Sum_Bound_2} but only carried out for $1\leq s\leq 10$. 

Since $\hat d<3\dfree$, it is not necessary to evaluate triple or higher order errors under this truncation. It is clearly seen from the figure that the best degree-$m$ CRC codes found by our procedure outperform the existing degree-$m$ codes for all $m$. Furthermore, their performance is either better than or similar to the existing degree-$(m+2)$ code except for $m=6$. In other words, the proposed design can typically save $2$ check bits while keeping the same error detection capability.

\begin{figure}[!t]
\centering
\ifCLASSOPTIONonecolumn
\includegraphics[width=0.53\columnwidth]{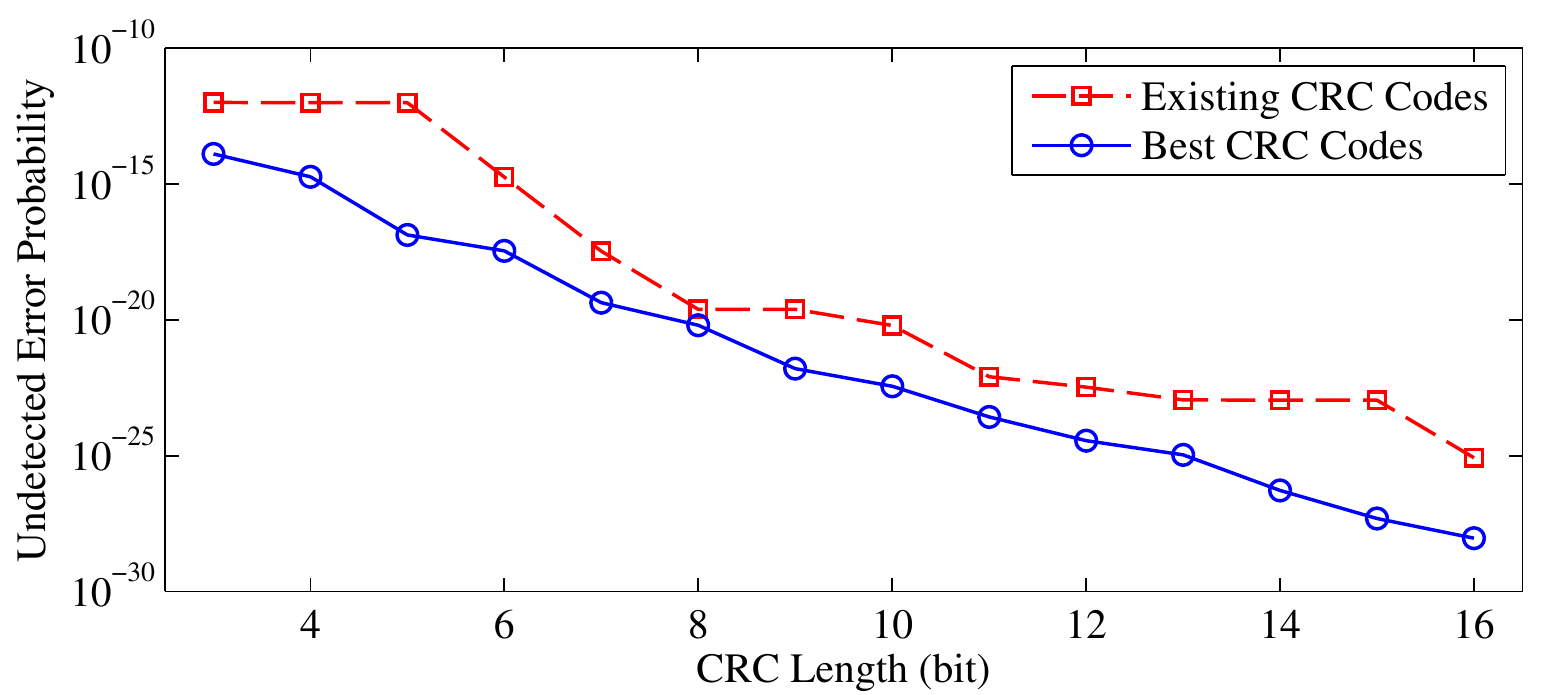}
\else
\includegraphics[width=0.99\columnwidth]{./figure/Undetected_FER_Gain}
\fi
\caption{The undetected error probability of the existing and best CRC codes for convolutional code $(133,171)_8$ with information length $k=1024$ bits at SNR = $8$ dB.}
\label{fig:Undetected_FER_vs_m}
\end{figure}

We can compare the error detection capability of all codes at a fixed SNR. If we draw a vertical line at SNR = $8$~dB on Fig.~\ref{fig:Undetected_FER_vs_SNR}, the intersections  can be plotted along with the associated CRC lengths $m$ in Fig.~\ref{fig:Undetected_FER_vs_m}.
The largest reduction of undetected error probability is about five orders of magnitude at $m=5$ where the existing CRC code has an undetected error probability of $1.01 \times 10^{-12}$ and the newly designed CRC code has an undetected error probability of $1.36 \times 10^{-17}$, based on the analysis. Note that these numbers are calculated at $8$ dB SNR, and the reduction gets more significant as SNR increases. 

Since the existing CRC codes are not tailored to the convolutional code, a higher degree code does not necessarily have a better performance. We observe that there are three almost horizontal region for the existing CRCs with degrees $m$ ranging from $3$ to $5$, from $8$ to $9$, and from $13$ to $15$. The reason is that the codes in each region have similar number of dominant undetectable errors as can be seen in Table~\ref{tbl:CRC_List}.  In contrast, the CRC codes designed using our procedure show steady improvement as the degree increases.

\begin{figure}[!t]
\centering
\ifCLASSOPTIONonecolumn
\includegraphics[width=0.53\columnwidth]{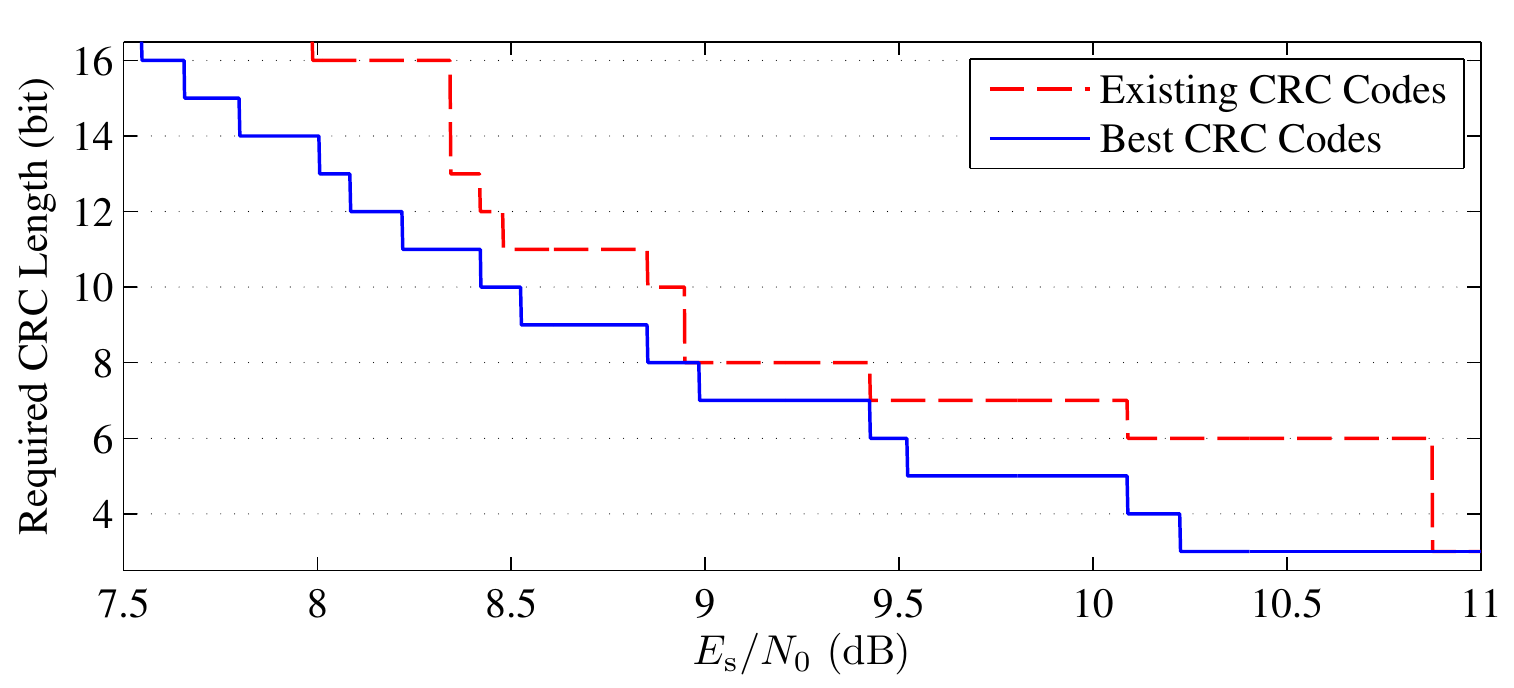}
\else
\includegraphics[width=0.99\columnwidth]{./figure/Required_CRC_Length}
\fi
\caption{The required CRC lengths for the existing and best CRC codes for convolutional code $(133,171)_8$ with information length $k=1024$ bits to achieve undetected error probability below $10^{-25}$.}
\label{fig:m_vs_SNR}
\end{figure}

The existing and best CRC codes can also be compared in terms of the required CRC length to achieve a certain undetected error probability. Assume our target is to reach undetected error probability below $10^{-25}$. This can be shown by drawing a horizontal line at the target probability level on Fig.~\ref{fig:Undetected_FER_vs_SNR} and plotting the CRC lengths associated to the crossed points as a function of SNR as in Fig.~\ref{fig:m_vs_SNR}.
For most of the SNR levels, the best CRC codes requires two fewer check bits than the existing CRC codes to achieve the same error detection capability. At SNR around $10.5$ dB, the best CRC code requires three fewer check bits than the existing CRC code. Since the existing code required six check bits, this is a 50\% reduction.

% [CONCLUSION]
\section{Conclusion}
\label{sec:Conclusion}
A good CRC code in a convolutionally coded system should minimize the undetected error probability. To calculate this probability, two methods based on distance spectrum are proposed. The exclusion method starts with all possible single and multiple error patterns of the convolutional code and excludes them one by one by testing if they are detectable. In the construction method, undetectable errors are mapped to error events of an equivalent convolutional code, which is the combination of the CRC code and the original convolutional code. The computer search for error events in the construction method does not need to record the error patterns and thus can go deeper than the search in the exclusion method. However, the construction method could encounter difficulties while dealing with high-degree CRC codes. Moreover, the construction method is generally applicable to the performance analysis of catastrophic convolutional codes.

We also propose a search procedure to identify the best CRC codes for a specified convolutional encoder and information length.  A candidate CRC code is excluded if it has more low-distance undetectable errors. Therefore, the best CRC polynomial is guaranteed to have the fewest dominant undetectable errors and minimizes the probability of undetected error when SNR is high enough. When undetectable double errors dominate, the choice of the best CRC polynomial is more dependent on the information length. In an example application of the design procedure for the popular $64$-state convolutional code with information length $k=1024$, new CRC codes provided significant reduction in undetected error probability compared to the existing CRC codes with the same degrees. With the proposed design, we are able to save two check bits in most cases while having the same error detection capability.

It is an open problem to generalize this work to other error-correcting codes, such as turbo or low-density parity-check (LDPC) codes. The construction method can only be applied when the encoder has the convolutional structure. We note that convolutional LDPC codes have such a structure. For turbo codes, the construction method can be used to analyze one of the recursive systematic convolutional (RSC) codes but analyzing the other RSC code is not straightforward due to the existence of the interleaver. Nevertheless, the exclusion method can be generalized to any other error-correcting codes if the dominant error events are identified.

% [APPENDIX]
\appendices
\ifCLASSOPTIONonecolumn
\section{Efficient Search for Undetectable Double Errors}
\else
\section{Efficient Search for Undetectable Double Errors}
\fi
\label{sec:Appendix_Search_Double_Error}
In \eqref{eqn:PUD_Double_Bound1} and \eqref{eqn:PUD_Double_Bound2}, the indicator function is evaluated for every $g_1$ given $e_1(x)$ and $e_2(x)$. However, there is a more efficient way to check the divisibility and can save a lot of computation time when $n+\nu$ is large.

The remainder of any polynomial in $\GF(2)[x]$ divided by $p(x)$ forms a quotient ring $\GF(2)[x]\slash p(x)$. Two polynomials mapped to the same element of the quotient ring are called \emph{congruent}. If $x^{g_1+l_2}e_1(x)$ is congruent to $e_2(x)$ modulo $p(x)$, then their combination forms an undetectable double error. To characterize the remainder of $x^{g_1+l_2}e_1(x)$ modulo $p(x)$, we apply the concept of cyclotomic coset\cite{Cyclotomic_Coset}, which is originally defined for integers, to polynomials. For polynomials in $\GF(2)[x]$, define $x$-cyclotomic coset modulo $p(x)$ containing $e(x)$ as 
\begin{equation}
\mathbf{C}_{e(x)} = \left\{\left. x^h e(x)\left(\bmod\,\,p(x)\right) \right|h = 0,1, \cdots \right\},
\label{eqn:Cyclotomic_Coset_Definition}
\end{equation}
which includes the remainder of all possible offsets of $e(x)$ divided by $p(x)$. One can verify that two cyclotomic cosets are either identical or disjoint, so all of the distinct $x$-cyclotomic cosets modulo $p(x)$ form a partition of quotient ring $\GF(2)[x]\slash p(x)$. For example, consider a degree-$2$ primitive polynomial $p(x)=x^2+x+1$, and we have $\mathbf{C}_0 = \left\{0 \right\}$ and $\mathbf{C}_1 = \left\{1,x,x+1 \right\}$ forming a partition of $\GF(2)[x]\slash p(x)$; consider a degree-$2$ non-primitive polynomial $p(x)=x^2+1$, and we have $\mathbf{C}_0 = \left\{0 \right\}$, $\mathbf{C}_1 = \left\{1,x \right\}$, and $\mathbf{C}_{x+1} = \left\{x+1 \right\}$ forming a partition of $\GF(2)[x]\slash p(x)$. In both cases, $\mathbf{C}_0$ is trivial since it only contains the ``zero'' element. In the primitive case, $\mathbf{C}_1$ is the only non-trivial cyclotomic coset and its cardinality is $\left|\mathbf{C}_1\right|=\left|\GF(2)[x]\slash p(x)\right| - 1=2^m-1$. In the non-primitive case, there are multiple non-trivial cyclotomic cosets and their sizes are smaller than $2^m-1$. In fact, there is only one unique non-trivial cyclotomic coset if $p(x)$ is a primitive polynomial.
% The complete proof is provided in Appendix~\ref{sec:Appendix_Primitive_Polynomial_Proof}.

It is obvious that if $e_1(x)$ and $e_2(x)$ belong to different cyclotomic cosets, there is no way to have a $g_1$ that makes $x^{g_1+l_2}e_1(x)$ congruent to $e_2(x)$ modulo $p(x)$. In other words, it is unnecessary to check whether any $g_1$ creates an undetectable double error with the specific $e_1(x)$ and $e_2(x)$. If $e_1(x)$ and $e_2(x)$ belong to the same cyclotomic coset $\mathbf{C}_{e_1(x)}$, only one proper $g_1\in \left[0,\left|\mathbf{C}_{e_1(x)}\right|-1\right]$ can make $x^{g_1+l_2}e_1(x) + e_2(x)$ divisible by $p(x)$. Denote this particular $g_1$ as $g'_1$. Once we find $g'_1$, all possible $g_1$ that create undetectable double errors are just $g_1 = g'_1 + u\left|\mathbf{C}_{e_1(x)}\right|$ for non-negative integer $u$ satisfying $g_1+l_1+l_2\leq n+\nu$.

Note that when $e_1(x) = e_2(x)$, they will belong to the same cyclotomic coset no matter what CRC generator polynomial $p(x)$ is. That is to say, no CRC code is able to detect such double error if these two error events have a proper gap $g_1$. Fortunately, the smallest proper gap is $g'_1=-l_2\left(\bmod\,\,\left|\mathbf{C}_{e_1(x)}\right|\right)$ so the total length of the undetectable double error is $\left|\mathbf{C}_{e_1(x)}\right|+l_1$. Hence, when $n+\nu$ is small enough, such double error will never occur. On the other hand, when $n+\nu$ is large, $d_{\min}$, which is the shortest distance of the undetectable errors, will be upper bounded by $2\dfree$.

\section{The Relationship between $\mathbf{\Delta}_i$ and $\mathbf{C}_{e(x)}$}
\label{sec:Appendix_Subset_Cyclotomic_Coset}
Define the state of the equivalent code at time $n-g$ as $q'_g(x)$, which is a polynomial with maximum degree $m+\nu-1$ representing consecutive $m+\nu$ bits from the $x^{g+m+\nu-1}$ term to the $x^g$ term in $x^{m}q'(x)$ for $g\in \left[-\nu, n \right]$. The corresponding state of the original code is given by the coefficients of the terms from $x^{m+\nu-1}$ to $x^m$ in polynomial $q'_g(x)p(x)$. Let $q'_{g,u}$ and $p_u$ be the coefficients of $x^u$ in $q'_g(x)$ and $p(x)$, respectively. Then the coefficient of $x^u$ for $u\in \left[m,m+\nu-1\right]$ in $q'_g(x)p(x)$ is given by
\begin{equation}
\sum_{v=0}^m{q'_{g,u-v}\,p_v}
\label{eqn:State_Relationship}.
\end{equation}
%If the state of the original code is known and $q'_{g,v}$ is known for $v\in \left[\nu,m+\nu-1\right]$, $q'_{g,\nu-1}$ can be solved through \eqref{eqn:State_Relationship} when $u=m+\nu-1$ because $p_m=1$. Moreover, the rest of $q'_{g,v}$ for $v=\nu-2, \cdots, 1, 0$ can be solved one by one in the same way.

Assume that the all-zero codeword is sent, i.e. $q(x)=0$, and the trellis path enters a detectable-zero state $\textnormal{S}^\textnormal{D}_i$ at time $n-g$. Also, let $e(x)$ be a polynomial of degree smaller than or equal to $n-g-1$ representing the length-$(n-g)$ input sequence of the original convolutional encoder from the beginning to time $n-g$, and it is given by the coefficients from the $x^{n-1}$ term to the $x^{g}$ term in $q'(x)p(x)$. Since the state at time $n-g$ is $\textnormal{S}^\textnormal{D}_i$, $e(x)$ must be non-divisible by $p(x)$.

The remainder of $e(x)$ divided by $p(x)$ is given by
\begin{equation}
e(x)\,\left(\bmod\,\,p(x)\right) =\sum_{u=0}^{m-1}{x^u \sum_{v=u+1}^m {q'_{g,m+u-v} p_v}}
\label{eqn:Remainder_of_e1(x)_by_p(x)},
\end{equation}
which is totally governed by $q'_g(x)$, or $\textnormal{S}^\textnormal{D}_i$. If the remainder is known, the bits $q'_{g,v}$ for $v\in\left[0,m-1\right]$ can be solved uniquely through back substitution for $u=m-1, m-2, \cdots, 0$ because $p_m=1$. Furthermore, the whole polynomial $q'_g(x)$, or $\textnormal{S}^\textnormal{D}_i$, can be solved by letting \eqref{eqn:State_Relationship} equal to zero for $u=m+\nu-1, m+\nu-2, \cdots, m$ because the state of the original code is just $\nu$ zeros. Hence, the remainder of $e(x)$ divided by $p(x)$ determines a detectable-zero state $\textnormal{S}^\textnormal{D}_i$, and vice versa. Furthermore, each of the $2^m-1$ non-zero elements in $\GF(2)[x]\slash p(x)$ corresponds to a unique state in $\SD$.

To find all states in $\mathbf{\Delta}_i$, we can specify $q'_{g-h,0}$, the input bit to the equivalent encoder at time $n-g+h$, for $h=1,2,\cdots$ such that the input bits to the original convolutional encoder after time $n-g$ are all zeros and thus the following states are in $\SD$. By doing so, we know that the polynomials $q'_{g-h}(x)$ will represent the states in $\mathbf{\Delta}_i$. This procedure is finished when certain $q'_{g-h}(x)$ represents $\textnormal{S}^\textnormal{D}_i$ again. During this procedure, the corresponding input sequence to the original encoder from the beginning to time $n-g+h$ is simply $x^he(x)$ because the input bits after time $n-g$ are all zeros. By the definition given in \eqref{eqn:Cyclotomic_Coset_Definition}, the remainder of $x^he(x)$ divided by $p(x)$ is an element of $\mathbf{C}_{e(x)}$. In addition, we know that this remainder corresponds to a state in $\mathbf{\Delta}_i$. Therefore, $\mathbf{\Delta}_i$ and $\mathbf{C}_{e(x)}$ contain the same elements but just represented in different forms.

\section*{Acknowledgment}
The authors would like to thank Dr. A. R. Williamson with Communications Systems Laboratory (CSL) at University of California, Los Angeles (UCLA) for useful discussions and Mr. K. Vakilinia with CSL at UCLA for his kind help.
% [REFERENCE]
\ifCLASSOPTIONcaptionsoff
  \newpage
\fi
\bibliographystyle{IEEEtranTCOM}
\bibliography{IEEEabrv,CRC_Convolutional_bib}

% [BIOGRAPHY]
%\begin{biography}[{\includegraphics[width=1in,height=1.25in,clip,keepaspectratio]{mshell}}]{Michael Shell}


%\vfill


\end{document}